\documentclass{article}

\usepackage{microtype}
\usepackage{graphicx}
\usepackage{subfigure}
\usepackage{booktabs} 

\usepackage{hyperref}

\usepackage[accepted]{icml2024}

\usepackage{amsmath}
\usepackage{amssymb}
\usepackage{mathtools}
\usepackage{amsthm}
\usepackage{pifont}
\usepackage{amsfonts,bm}
\usepackage{multirow}
\usepackage{wrapfig}
\usepackage{colortbl}
\usepackage{xcolor}
\usepackage{xspace}
\usepackage{soul}

\usepackage[capitalize,noabbrev]{cleveref}

\theoremstyle{plain}

\theoremstyle{definition}

\theoremstyle{remark}

\usepackage[textsize=tiny]{todonotes}

\newcommand{\ie}{\textit{i}.\textit{e}.}
\newcommand{\eg}{\textit{e}.\textit{g}.}

\newcommand{\Fref}[1]{Figure~\ref{#1}}

\newcommand{\Name}{\texttt{AquaLoRA}\xspace}
\newcommand{\method}[0]{AquaLoRA}

\newcommand{\cmark}{\ding{51}}%
\newcommand{\xmark}{\ding{55}}%

\icmltitlerunning{Toward White-box Protection for Customized Stable Diffusion Models via Watermark LoRA}

\begin{document}

\twocolumn[
\icmltitle{\Name: Toward White-box Protection for Customized \\ Stable Diffusion Models via Watermark LoRA}



\icmlsetsymbol{equal}{*}
\icmlsetsymbol{cor}{\dag}

\begin{icmlauthorlist}
\icmlauthor{Weitao Feng}{equal,ustc}
\icmlauthor{Wenbo Zhou}{equal,ustc}
\icmlauthor{Jiyan He}{ustc}
\icmlauthor{Jie Zhang}{cor,ntu}
\icmlauthor{Tianyi Wei}{ustc}
\icmlauthor{Guanlin Li}{ntu}
\icmlauthor{Tianwei Zhang}{ntu}
\icmlauthor{Weiming Zhang}{ustc}
\icmlauthor{Nenghai Yu}{ustc}
\end{icmlauthorlist}

\icmlaffiliation{ustc}{University of Science and Technology of China, Hefei, China}
\icmlaffiliation{ntu}{Nanyang Technological University, Singapore}

\icmlcorrespondingauthor{Jie Zhang}{jie\_zhang@ntu.edu.sg}

\icmlkeywords{Machine Learning, ICML}

\vskip 0.3in
]



\printAffiliationsAndNotice{\icmlEqualContribution} 

\begin{abstract}

Diffusion models have achieved remarkable success in generating high-quality images. Recently, the open-source models represented by Stable Diffusion (SD) are thriving and are accessible for customization, giving rise to a vibrant community of creators and enthusiasts.
However, the widespread availability of customized SD models has led to copyright concerns, like unauthorized model distribution and unconsented commercial use. 
To address it, recent works aim to let SD models output watermarked content for post-hoc forensics. 
Unfortunately, none of them can achieve the challenging white-box protection, wherein the malicious user can easily remove or replace the watermarking module to fail the subsequent verification.
For this, we propose \texttt{\method} as the first implementation under this scenario. Briefly, we merge watermark information into the U-Net of Stable Diffusion Models via a watermark Low-Rank Adaptation (LoRA) module in a two-stage manner.
For watermark LoRA module, we devise a scaling matrix to achieve flexible message updates without retraining. 
To guarantee fidelity, we design Prior Preserving Fine-Tuning (PPFT) to ensure watermark learning with minimal impacts on model distribution, validated by proofs. Finally, we conduct extensive experiments and ablation studies to verify our design.
Our code is available at \href{https://github.com/Georgefwt/AquaLoRA}{github.com/Georgefwt/AquaLoRA}.

\end{abstract}
\section{Introduction}

\begin{figure*}[htb]
\vskip 0.1in
\begin{center}
\centerline{\includegraphics[width=0.95\textwidth]{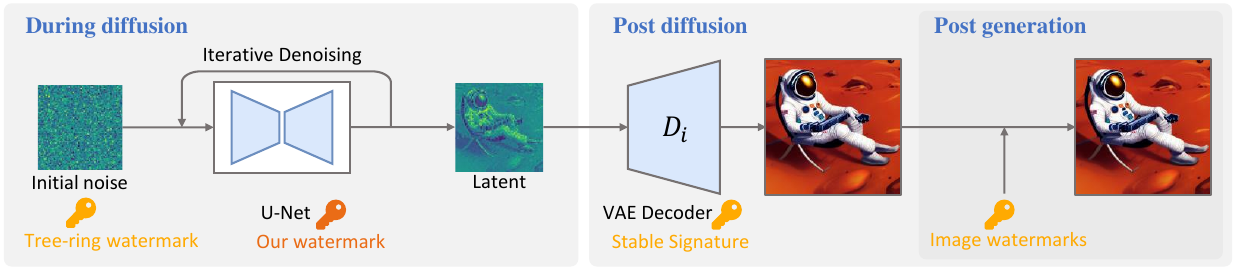}}
\caption{Illistration of different watermark placement with the Stable Diffusion model. Our watermark is embedded within the core structure of the diffusion model, the U-Net.}
\label{fig:method_comp_demo}
\end{center}
\vskip -0.2in
\end{figure*}

With the flourishing development of generative AI and cross-modal visual and language representation learning~\cite{radford2021learning,yuan2021florence}, text-to-image (T2I) synthesis models~\cite{ramesh2022hierarchical, saharia2022photorealistic, rombach2022high} have gained popularity due to their convenient interactions and high-fidelity synthetic results. 
As a standout in the realm of T2I models, the universe of Stable Diffusion is thriving, fueled by its complete open-source nature. Various versions of models (e.g., v1, v2, XL, etc.) and customized technologies~\cite{ruiz2023dreambooth,gal2022image} are constantly emerging, providing immense enjoyment and have fostered active communities (\eg, \citealp{civitai}, \citealp{prompthero}, \citealp{patreon}) where users can exchange or sell their customized Stable Diffusion models.

This ease of sharing raises copyright concerns, such as the unconsented use of generated images and redistribution of customized models for profit, potentially compromising the interest of original creators.
The official repository of Stable Diffusion models offers some ad-hoc image watermarking methods \cite{rahman2013dwt,zhang2019robust} as a makeshift protection. Afterward, additional efforts~\cite{wen2023tree,fernandez2023stable} propose \textbf{intergrated watermarking}, namely, integrating the watermarking process more intricately with the generation process, including factors like initial noise and the VAE decoder. All the above watermarking approaches are illustrated in \Fref{fig:method_comp_demo}.
In this paper, we consider a more challenging protection scenario, namely, \textbf{white-box protection}, wherein adversaries have full access to the watermarked SD models. Because SD models are wildly open-source, it's easy for adversaries to bypass watermarking by changing the sampling strategy or replacing the VAE, making all current watermarking protection ineffective.

To remedy it, we propose the insight that \textit{\ul{watermarking should be coupled with the most crucial component of Stable Diffusion}}.
Thus, we suggest embedding the watermark directly into U-Net, the most central structure containing essential knowledge. In such a mechanism, the disruption of watermarking is accompanied by a significant drop in generation fidelity. 
Besides, there are three additional requirements: 1) \textbf{Fidelity:} high visual quality between the watermarked generated image and the watermark-free one,
2) \textbf{Robustness:} the watermark shall be robust against different image distortions and generation configurations, and 3) \textbf{Flexibility:} for large-scale and multi-user deployments, it is essential to have a large watermark capacity, while ensuring that the embedding and extraction processes for each user do not incur additional training overhead.

To satisfy the above requirements, we propose \Name, a two-stage watermarking framework, consisting of latent watermark pre-training and watermark learning with prior preserving.
In the first stage, we transfer the philosophy of image watermarking to the latent space of Stable Diffusion to create a watermark pattern suitable for the U-Net to learn. We thoroughly consider the robustness of the watermark and propose the Peak Regional Variation Loss (PRVL) to enhance fidelity further. The trained secret encoder is private as a confidential codebook.
In the second stage, we introduce a prior preserving fine-tuning (PPFT) method that allows the watermark pattern from the previous stage to be learned by U-Net, while minimally perturbing the original knowledge of the model.
To achieve integrated watermarking, we propose Watermark LoRA, which represents watermark information by a scaling matrix, and merge it into the original model weights so that it cannot be easily removed.
Moreover, Watermark LoRA is trained with different watermark information, inherently satisfying flexibility requirements.
Finally, we adopt a coarse type adaption to enhance performance further. The proposed \Name shows good adaptability on different customized Stable Diffusion models.

\vspace{-0.5em}
Our contributions can be summarized as follows:
\begin{itemize}
    \item We point out the necessity for white-box protection to Stable Diffusion model and propose \texttt{\method} as the first implementation of a white-box protection watermark for current customized Stable Diffusion models.
    \item We apply a two-stage design. The distortion layer in the first stage guarantees the robustness of our method; the proposed scaling matrix for Watermark LoRA module strategy grants our scheme's flexibility; more importantly, the well-devised prior preserving fine-tuning method and PRVL substantially enhances the fidelity.
    \item Sufficient experiments and ablations prove that our watermark meets all the previously mentioned requirements as well as the effectiveness of proposed designs.    
\end{itemize}

\section{Related Work}

\subsection{Stable Diffusion Models}

Stable Diffusion has gone viral due to its powerful generative capabilities and its open-source nature. The overall pipeline of Stable Diffusion follows that of latent diffusion \cite{rombach2022high}, mapping images into latent space and performing diffusion in the latent space of a VAE, which significantly reduces the computation cost. It can be considered a representative example of latent diffusion. Many novel application scenarios have emerged from Stable Diffusion, where customization is a key factor. People can make the model ``learn" entirely new styles and characters, enabling personalized generation.

Textual Inversion \cite{gal2022image}, one of the earliest methods in this field, can be considered a form of prompt optimization. Users represent the target content with a special token and continuously optimize the embedding of this token using target images.

Fine-tuning, as compared to prompt optimization, enables the most extensive customization of various generated content by adjusting the entire model. Dreambooth \cite{ruiz2023dreambooth} is one of the fine-tuning techniques that involves a unique way for a diffusion model to learn a special subject using a small number of specific images.
With increasing understanding and recognition of fine-tuning diffusion models, fine-tuning methods are being increasingly used to introduce more preferences into diffusion models, altering the model's style and even domain. LoRA, originally designed for fine-tuning Large Language Models, has proven effective for diffusion model fine-tuning as well.
It's important to note that LoRA is essentially a fine-tuning method and is not limited to personalized generation. In our work, we use LoRA to learn watermark patterns for watermarking.

\subsection{Watermarking Generative Models} \label{sec:GMwatermark}
With the popularity of generative models, there is a growing recognition of the importance of adding watermarks to AI-generated content or the corresponding generative models. 
The simplest makeshift is directly watermarking generated images.
Especially, Stable Diffusion \citeauthor{sd} suggests watermarking techniques like DWT-DCT, DWT-DCT-SVD \cite{rahman2013dwt}, and RivaGAN \cite{zhang2019robust}.  Unfortunately, removing the post-generation watermark can be achieved by merely altering a few lines of code.
Some approaches \cite{yu2021artificial,zhao2023recipe} suggest embedding watermarks in the entire training set, resulting in the generated models being able to effectively incorporate watermarks into all images they generate. However, for large-scale diffusion models, this approach is infeasible, as these large models are trained on massive datasets.

Afterward, some works try to integrate the watermarking process with the generation process. For example, Stable Signature \cite{fernandez2023stable}, focusing on the Variational Autoencoder \cite{kingma2013auto} in Stable Diffusion Models, embeds watermarks in the VAE decoder, showing strong performance and removing the need for post-generation watermarking.
Another study \cite{xiong2023flexible} also modifies the VAE decoder structure by introducing a ``Message Matrix'', allowing for easy message updates without re-training the model. However, both methods are vulnerable in white-box scenarios, especially if a clean VAE decoder is available publicly.
Additionally, Tree-ring \cite{wen2023tree} targets the sampling process of the Diffusion model. They skillfully utilize the inherent properties of the diffusion model, placing a watermark pattern on the model's initial noise. It requires using a deterministic sampler, such as DDIM \cite{song2020denoising}, during image generation. This method uses DDIM inversion for watermark extraction, detecting watermarks by reverting images to their initial noise. The downside is that it requires the model owner to control the model users' sampling process, typically through an API. Similarly, in a white-box scenario, users can control the model's sampling process, making this method ineffective.

As shown in Figure \ref{fig:method_comp_demo}, our method adds the watermark to U-Net, utilizing its uniqueness to achieve watermarking in white-box scenarios.

\section{Preliminaries}

\begin{figure*}[t]
\vskip 0.1in
\begin{center}
\centerline{\includegraphics[width=0.96\textwidth]{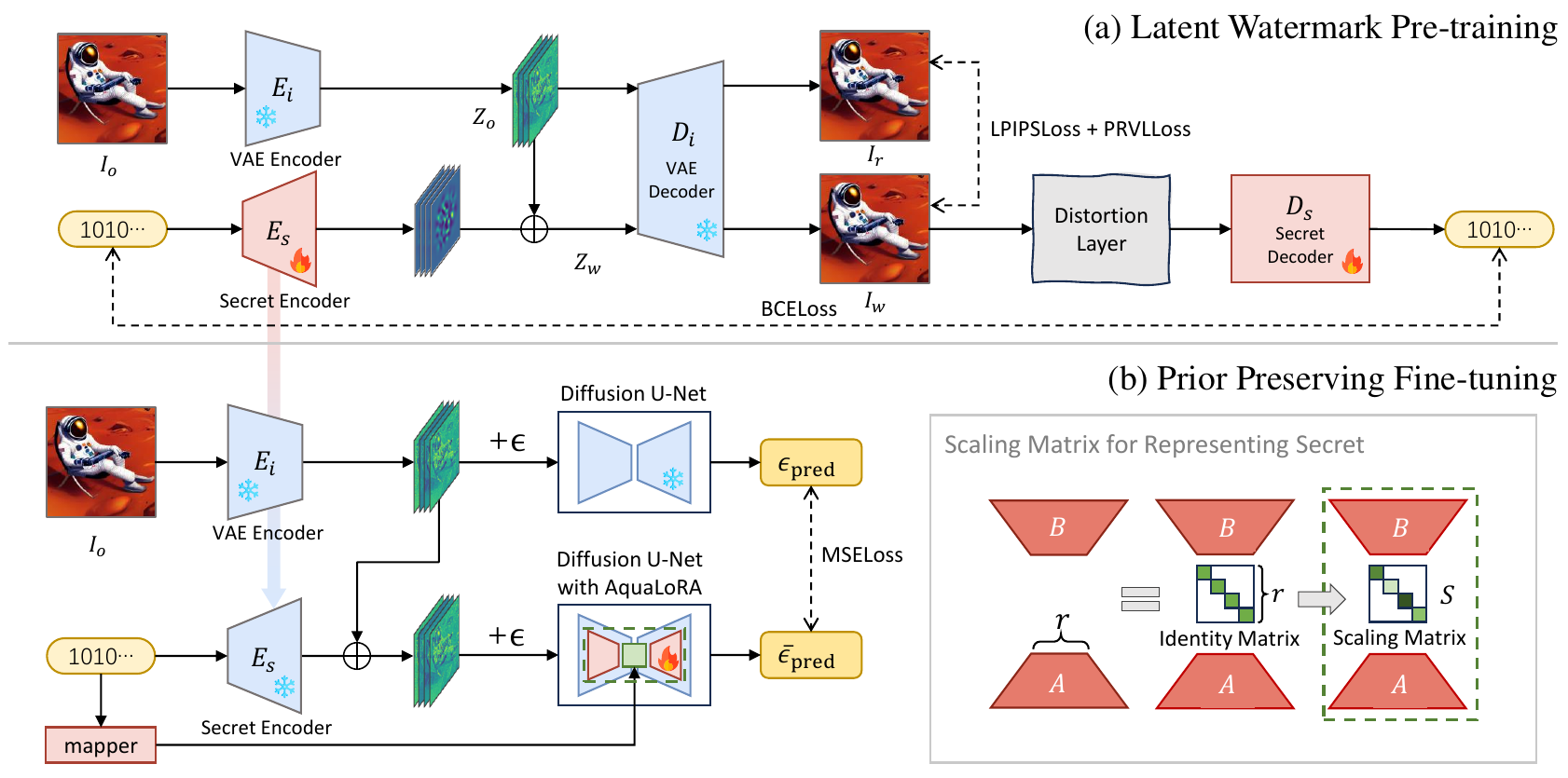}}
\caption{The overall framework of our method. (a) The first stage is latent watermark pre-training. In this phase, we jointly train a watermark secret encoder $E_s$ and decoder $D_s$ at the latent level. (b) After latent watermark pre-training, we employ our proposed prior preserving fine-tuning (PPFT) strategy to train AquaLoRA, which can be merged into any fine-tuned model weights, offering protection. Coarse type adaptation is omitted here, as it follows the same PPFT strategy.}
\label{fig:framwork}
\end{center}
\vskip -0.2in
\end{figure*}

\noindent \textbf{Latent Diffusion Model.}
LDMs incorporate a conditional denoising model, represented as \( \bm{\epsilon}_{\theta}(z_t, t, c) \), which is capable of generating images conditioned on a specific text \( c \). \( z_t \) denotes the latent representation at a specific timestep \( t \) within the range of \( \{1,...,T\} \).

During the training stage, a loss $\mathcal{L}_{\text{simple}}$ is leveraged to compel LDM to denoise the latent representations $z_t := \sqrt{\bar\alpha_t} z_0 + \sqrt{1 - \bar\alpha_t} \bm{\epsilon}$ as follows:
\begin{align}
\mathcal{L}_{\text{simple}} = \mathbb{E}_{z_0, {\bm{\epsilon}}, t, c}\left[\left\|\bm{\epsilon}_{\theta}\left(z_{t}, t, c\right)-\bm{\epsilon} \right\|_{2}^{2}\right], \label{eq:diffusion-training}
\end{align}
where $\alpha_t$ represent the parameters of the diffusion process, $\bm{\epsilon}$ is sampled from the Gaussian distribution $\mathcal{N}(\mathbf{0}, \mathbf{I})$, and $\bm{\epsilon}_{\theta}(z_t, t, c)$ is implemented as a text-conditional U-Net.

\noindent \textbf{Low-Rank Adaption.} LoRA is a method designed for efficiently adapting large-scale language and vision models to new tasks\cite{hu2021lora}. The key principle of LoRA is that the weight updates, denoted as $\Delta \mathbf{W}$, to the original model weights $\mathbf{W} \in \mathbb{R}^{n \times m}$ during fine-tuning exhibit a low intrinsic rank. Consequently, the update $\Delta \mathbf{W}$ can be represented as the product of two low-rank matrices $\mathbf{A} \in \mathbb{R}^{m \times r}$ and $\mathbf{B} \in \mathbb{R}^{r \times n}$, where $ \Delta \mathbf{W}$ is computed as $\mathbf{A} \times \mathbf{B}$.
In the training process, only the matrices $\mathbf{A}$ and $\mathbf{B}$ are updated, while original weights $\mathbf{W}$ remain unchanged.

During the inference, the forward computation is represented by $\mathbf{W}x + \mathbf{A} \mathbf{B}x$, $x$ is the output of the former layer in the neural network.
LoRA can seamlessly integrate into the original model using the formula $\mathbf{W}_{\text{updated}} = \mathbf{W} + \alpha \cdot \mathbf{A} \mathbf{B}$, with $\alpha$ usually set as 1.

In this paper, the proposed \ul{AquaLoRA (i.e., watermark LoRA)} is specifically designed for integrating watermark information with the target U-Net module.

\section{\method}

\subsection{Overview}

Figure \ref{fig:framwork} provides an overview of our method. Our approach is generally divided into two stages: latent watermark pertaining and prior preserving fine-tuning. The purpose of the latent watermark pre-training stage is to train a latent watermark scheme as a sort of codebook. In the prior preserving fine-tuning stage, this latent watermark pattern is learned by our proposed AquaLoRA through fine-tuning, which can be easily integrated into the model weights. 
For practical application, we can fine-tune AquaLoRA on checkpoints of various coarse types to create domain-specific versions, which can further boost performance.

\subsection{Latent Watermark Pre-training}

\noindent\textbf{Watermark Scheme Design.} In this stage, we aim to train an image watermark that is easily learned by the U-Net model. 
To achieve this goal, we first examine the challenges posed by existing image watermarks when it comes to diffusion models. We have identified two primary reasons for these challenges:
1) The watermark information tends to be disrupted or even lost when it is transformed into the latent space by the VAE encoder. This makes it extremely difficult for the diffusion model to effectively learn the watermark.
This can be verified by calculating the extraction accuracy of the watermark from the image reconstructed by the VAE. 2) previous work \cite{cui2023diffusionshield} introduced the concept of pattern uniformity, defined as the consistency of watermarks injected into different samples.
It has been observed that the higher the watermark's consistency, the more conducive it is for the watermark to be learned effectively. 
For a cover-agnostic watermark, the consistency is naturally the highest.
Thus, we set our design goal: a cover-agnostic watermark that is prominent in the latent space.

\noindent\textbf{Training Pipeline Design.} 
As analyzed above, we aim to train a cover-agnostic watermark that is prominent in the latent space. As shown in Figure \ref{fig:framwork}(a), in this stage, the trained secret encoder $E_s$ will be utilized in the next training stage, acting as a sort of codebook that is secured, while the secret decoder $D_s$ will serve as the final watermark extractor. The process of watermark embedding and extraction adheres to the conventional encoder-decoder structure, but with the distinction that the watermark is injected in the latent space of a VAE. Here, we adopt a simple addition operation as the watermarking embedding process: 
\begin{eqnarray}
    I_w = D_i(E_s(s) + E_i(I_o)),
\end{eqnarray}
where $E_i$ and $D_i$ are the VAE encoder and decoder of the latent diffusion, respectively. $I_o$ is the original image and $s$ is secret (i.e., watermark). Architecture for secret encoder $E_s$ is inspired by \cite{bui2023rosteals}, and more details can be found in Appendix \ref{app:secret encoder_train strategy}.
For the secret decoder, we adopt EfficientNet-B1\cite{tan2019efficientnet} to directly retrieve watermark information from the watermarked image, denoted as $I_w$. This process involves computing the Binary Cross-Entropy Loss (BCELoss) with the original information.

To ensure visual consistency, we calculate the LPIPS loss \cite{zhang2018unreasonable} between the watermarked image \(I_w\) and reconstructed image \(I_r\). We do not calculate it between \(I_w\) and \(I_o\) because the image already suffers from quality loss due to compression by the VAE encoder and reconstruction by the decoder. We do not expect the watermark to learn image restoration, as it would increase the training difficulty. Moreover, to reduce artifacts produced by the watermark and enhance fidelity, we designed the Peak Regional Variation Loss (PRVL). The detailed design of PRVL can be found in the Appendix \ref{app:PRVL detail}.

Overall, our training objective can be summarized below, where $\lambda$ and $\mu$ are coefficients:
\begin{align}
\mathcal{L}_{\mathrm{Total}} = \mathcal{L}_{\mathrm{BCE}} + \lambda \mathcal{L}_{\mathrm{LPIPS}} + \mu\mathcal{L}_{\mathrm{PRVL}}.
\end{align}

\subsection{Watermark Learning with Prior Preserving}

This stage integrates the previously generated watermark pattern into U-Net.
To accomplish this, we leverage the remarkable adaptability and straightforward integration capabilities of LoRA, thanks to its minimal perturbation of the model's prior settings.

\subsubsection{Scaling Matrix for Representing Secret}

To achieve flexibility, allowing arbitrary changes to the embedded secret during utilization, we need to add a structure to LoRA that introduces a new condition on secret $\mathbf{s}$. 
To this end, we modify the structure of LoRA by introducing a \textbf{scaling matrix}. The computation formula for LoRA can be then written as $ \Delta \mathbf{W} = \mathbf{A} \times \mathbf{S} \times \mathbf{B}$, where $ \mathbf{A} \in \mathbb{R}^{n \times r} $ and $ \mathbf{B} \in \mathbb{R}^{r \times m} $, $ \mathbf{S} \in \mathbb{R}^{r \times r}$ as the scaling matrix. When $\mathbf{S}$ is an identity matrix, the computation yields the same result. As shown in Figure \ref{fig:framwork}(b), by relaxing the constraints, we create space to introduce a secret message into LoRA.

Then, we explored the use of learnable embeddings to design a mapper that transforms a secret of length $l$ into a vector of length $r$. 
Specifically, for the $i$-th bit of a secret, we use a embedding vector $ \mathbf{I}_i $ and $ \mathbf{0} $ to represent the binary states 1 and 0, with $ \mathbf{I}_i, \mathbf{0} \in \mathbb{R}^r $. The mapping function $ f_i: \{0, 1\} \rightarrow \mathbb{R}^r $ is then defined by
\begin{equation}
    f_i(b_i) = 
    \begin{cases}
        \mathbf{I}_i, & \text{if } b_i = 1, \\
        \mathbf{0}, & \text{otherwise}.
    \end{cases}
\end{equation}
For a given secret $ s = \{b_0, b_1, \ldots, b_l\} $, the scaling matrix $ \mathbf{S} $ is constructed as 
\begin{equation}
    \mathbf{S} = \mathrm{diag}\left( \textbf{1} + \frac{1}{\sqrt{l}} \sum_{i=1}^{l} f_i(b_i) \right).
\end{equation}
In each iteration of the training, we use a batch of random secrets for the forward pass. In application, when adding a watermark to the model, for a target secret, we pass the secret through the aforementioned mapper to obtain the scaling matrix $\mathbf{S}$. We then calculate the final $\Delta \mathbf{W}$ and merge it into the model weights by calculating $\mathbf{W}_{\text{watermarked}} = \mathbf{W} + \alpha \Delta\mathbf{W}$.

In the default setting, $\mathbf{I}_i$ is initialized using a standard normal distribution. Considering that it is beneficial to maximize the difference between vectors, we propose using orthogonal initialization. That is, for any $ \mathbf{I}_j, \mathbf{I}_k, j,k \in [0,l]$, we have $ \mathbf{I}_j \cdot \mathbf{I}_k = \textbf{0}$. Experiments show that this leads to further improvement in performance. The comparison results of these initialization methods can be found in Table \ref{tab:ablation}.

\subsubsection{Prior Preserving Fine-tuning}

Here, we introduce our specially designed fine-tuning method to ensure fidelity while learning the watermark into AquaLoRA. For a fixed secret, our watermark is cover-agnostic and can be formalized as a specific fixed offset \(\Delta z_w\) to the distribution. The most naive approach to teaching a Diffusion Model to learn a fixed offset would be to find an Image-Caption dataset and simply use the Diffusion Model’s training loss (Equation \ref{eq:diffusion-training}). However, this method has a serious issue: the data distribution of the Diffusion Model uncontrollably shifts closer to that of the Image-Caption dataset during training, resulting in significant changes in the generated outputs.

To address this issue, we analyze this problem and propose prior preserving fine-tuning (PPFT), as illustrated in Figure~\ref{fig:framwork}(b), which solves this issue well.
Let us denote the noise prediction result of the model at input timestep $t$ as $\boldsymbol\epsilon_{\mathrm{pred}}$. The corresponding equation can be formalized as:
$    \boldsymbol{\epsilon}_{\vartheta}(z_t, t, c) = \boldsymbol\epsilon_{\mathrm{pred}}.
$
Following the Denoising Diffusion Probabilistic Model (DDPM) \cite{ho2020denoising}, we express $z_t$ as:
$z_t = \sqrt{\bar{\alpha}_{t}} z_{0} + \sqrt{1-\bar{\alpha}_{t}} \boldsymbol{\epsilon}.
$

The learning target of DDPM is $\boldsymbol{\epsilon}$. By setting the learning target as $\boldsymbol{\epsilon}$, the model will try to fit the distribution of the training data, which directly leads to a distributional shift.
The target distribution we pursue is not the distribution of the dataset, but rather the intrinsic distribution of the model itself, augmented by an offset. Therefore, the best estimation should come from the predictions of the original model \({\boldsymbol\epsilon}_{\mathrm{pred}}\), rather than the actual added noise. Based on this observation, we begin our derivation with the evidence lower bound (ELBO), advancing towards the formulation of the final expression of $\mathcal{L}_{\mathrm{PPFT}}$. For a comprehensive understanding of this derivation process, please refer to the detailed explanation provided in Appendix \ref{app:PPFT-math}.

Finally, our Distribution Preservation Simpleloss can be formalized as:
\begin{equation}
\begin{split}
    \mathcal{L}_{\mathrm{PPFT}}(\theta) := \mathbb{E}_{t, c, z_{0}, \boldsymbol{\epsilon}}\Big[&
    \left\| \boldsymbol{\epsilon}_{\theta}\left(\sqrt{\bar{\alpha}_{t}} (z_{0} + \Delta z_w)\right.\right.\\
    &\hspace{-8em}\left.\left. + \sqrt{1-\bar{\alpha}_{t}} \boldsymbol{\epsilon}, t, c\right) - \boldsymbol{\epsilon}_{\vartheta}\left(\sqrt{\bar{\alpha}_{t}} z_{0} + \sqrt{1-\bar{\alpha}_{t}} \boldsymbol{\epsilon}, t, c\right)
    \right\|^2 \Big],
\end{split}
\end{equation}
where \(\theta\) represents the parameters of the fine-tuned model, \(\vartheta\) denotes the parameters of the original model, which are frozen, and $z_{0}$ is the latent of the image from the training dataset.

The pseudo-code of prior preserving diffusion fine-tuning is presented in Algorithm \ref{alg:loratuning}.

\begin{algorithm}[t]
\caption{Prior Preserving Fine-tuning Algorithm} \label{alg:loratuning}
\begin{algorithmic}[1]
    \STATE \textbf{Input:} Pre-trained frozen model $\vartheta$, AquaLoRA $\Delta \theta$, Pre-trained secret encoder $E_s$, diffusion model VAE encoder $E_i$. An image-caption dataset with paired images and captions.
    \STATE \textbf{Output:} Fine-tuned AquaLoRA $\Delta \theta$
    
    \FOR{image $\mathbf{x}_0, c$ in Dataset}
        \STATE $z_0 \gets E_i (\mathbf{x}_0)$
        \STATE $\mathbf{s} \gets \text{random secret}$
        \STATE $\theta(\mathbf{s}) \gets \vartheta + \Delta \theta(\mathbf{s})$
        \STATE $\Delta z_w \gets E_s(\mathbf{s})$
        \STATE $t \sim \operatorname{Uniform}(\{1, \ldots, T\})$
        \STATE $\boldsymbol\epsilon \sim \mathcal{N}(\mathbf{0}, \mathbf{I})$
        \STATE $z_t \gets \sqrt{\bar{\alpha}_{t}} z_{0} + \sqrt{1-\bar{\alpha}_{t}} \boldsymbol{\epsilon}$
        \STATE $z^{\mathrm{ub}}_t \gets \sqrt{\bar{\alpha}_{t}} (z_{0} + \Delta z_w) + \sqrt{1-\bar{\alpha}_{t}} \boldsymbol{\epsilon}$
        \STATE $\text{Take gradient descent step on}$ \\ \qquad $\nabla_{\Delta\theta} \left\| \boldsymbol{\epsilon}_{\theta(\mathbf{s})}\left(z^{\mathrm{ub}}_t, t,c\right) - \boldsymbol{\epsilon}_{\vartheta}\left(z_t, t,c\right) \right\|^2$
    \ENDFOR
    \STATE \textbf{return} $\Delta \theta$
\end{algorithmic}
\end{algorithm}

\begin{table*}[t]
\caption{Comparison between our method and previous watermarking methods. The capacities of DwtDctSvd, RivaGAN, StableSignature, and our method are 64bit, 32bit, 48bit, and 48bit, respectively. We control the FPR at $10^{-6}$ and evaluate the TPR. As Tree-ring is a zero-bit watermark, the bit accuracy can't be calculated here. Adv. (Adversarial) here refers to the average performance when images are under different distortions. The top-2 results of the robustness metrics have been emphasized.}
\label{tab:main_res}
\vskip 0.1in
\begin{center}
\begin{scriptsize}
\begin{sc}
\setlength\tabcolsep{4.7pt}
\begin{tabular}{lccccccccc} 
\toprule
\multirow{2}{*}{Method}     & \multirow{2}{*}{\begin{tabular}[c]{@{}c@{}}Integrated\\Watermarking\end{tabular}} & \multirow{2}{*}{\begin{tabular}[c]{@{}c@{}}Watermarking\\Flexibility\end{tabular}} & \multirow{2}{*}{\begin{tabular}[c]{@{}c@{}}White-box\\ protection\end{tabular}} & \multicolumn{2}{c}{Fidelity} & \multicolumn{4}{c}{Robustness}  \\ 
\cmidrule(l){5-6}\cmidrule(l){7-10}
  & &  &     & \enspace FID $\downarrow$ & \enspace DreamSim$\downarrow$ &  BitAcc.$\uparrow$ & BitAcc.(Adv.)$\uparrow$ & TPR $\uparrow$ & TPR (Adv.) $\uparrow$  \\ 
\midrule
None & --    & --     & --   & 24.26 & -- & -- & -- & -- & --   \\ 
\midrule
\multicolumn{10}{l}{\textit{Post-diffusion}} \\
DwtDctSvd      & \xmark & \cmark  & \xmark & 23.84 & 0.017    & \textbf{100.0} &  70.55       & \textbf{1.00}  & 0.356    \\
RivaGAN        & \xmark & \cmark  & \xmark & 23.26 & 0.023    & \textbf{98.78}  & 84.19       & 0.983  &  0.630   \\
StableSig.     & \cmark & \xmark  & \xmark & 24.77 & 0.018    & 98.30    &  77.01     & 0.993  &  0.580   \\ 
\midrule
\multicolumn{10}{l}{\textit{During diffusion}}    \\
Tree-ring      & \cmark & \cmark  & \xmark & 24.91            & 0.301          & -- & --       & \textbf{1.00}  & 0.810    \\
$\text{Ours}_\text{SD}$    & \cmark & \cmark  & \cmark & 24.88 & 0.201    & 95.79    &  \textbf{91.86}       & 0.990  & \textbf{0.906}    \\
$\text{Ours}_\text{CustomAvg}$ & \cmark & \cmark  & \cmark & -- & 0.204  & 94.81  &  \textbf{90.27}      & 0.976  &   \textbf{0.861}  \\
\bottomrule
\end{tabular}
\end{sc}
\end{scriptsize}
\end{center}
\vskip -0.1in
\end{table*}

\subsubsection{Coarse Type Adaption for Diverse Models}

We aim to safeguard customized Stable Diffusion Models, which may deviate from the original Stable Diffusion v1.5 model in terms of their distribution. 
Notably, certain models, such as those designed in an anime-cartoon style, exhibit significant distribution disparities that can lead to a decline in performance.
To address this issue, we propose a straightforward yet highly effective solution: fine-tuning our AquaLoRA on various coarse types. This approach allows us to create specialized AquaLoRAs tailored for different model types, thereby minimizing the distribution gap.
These types don't need to be very specific, and all types can be seen in Appendix \ref{tab:evalmodels}. In application, our goal is to select the AquaLoRA with the closest distribution, and we simply use the corresponding coarse type to measure the gap. This strategy effectively enhances performance. The effectiveness can be found in the ablation section \ref{sec:ablation}.

\section{Experiments}

\subsection{Experiment Setup}

\noindent \textbf{Datasets.} During the latent watermark pre-training process, we use the COCO2017 \cite{lin2014microsoft} dataset and randomly select 10,000 images from the training set to train the latent watermark. 
In the Prior Preserving Fine-tuning stage, to better avoid distribution shifts, we leverage captions for 10,000 images from the COCO train set used before, along with 10,000 prompts from Stable-Diffusion-Prompts \cite{Gustavosta}. Besides, the generation process employs the dpmsolver \cite{lu2022dpm} multistep scheduler, sampling in 30 steps and a default guidance scale of 7.5, to generate corresponding images as our training set.

\noindent \textbf{Implement Details.} We use Stable Diffusion v1.5 as the base model. The number of embedded bits we designed is 48 bits. During latent watermark pre-training stage, we set $\lambda=5, \mu=0.5$, and adopt the AdamW optimizer with a learning rate of $1 \times 10^{-3}$, weight decay $1 \times 10^{-4}$, training for 40 epochs. In this phase, we introduce a distortion layer for robustness enhancement. 
Details of the distortion layer can be found in Appendix \ref{app:distortion layer}. The training strategy is discussed in Appendix \ref{app:secret encoder_train strategy}.
In the PPFT stage, we use a LoRA with a rank of 320 by default as the base of our AquaLoRA. Our design generally follows the Kohya\_ss style \cite{kohya_ss}, including LoRA on the feedforward network in TransformerBlock and the conv layer in the ResBlock structure.
We also use the AdamW optimizer in this stage, with a learning rate of $1 \times 10^{-4}$, training for 30 epochs.

In the sampling phase, adjusting the $\alpha$ value allows for an easy trade-off between fidelity and watermark extraction accuracy. We choose $\alpha=1.05$, experiment can be found in the Appendix \ref{sec:tradeoff}.

\subsection{Fidelity}

\begin{figure}[t]
\vskip 0.1in
\centerline{\includegraphics[width=1.\columnwidth]{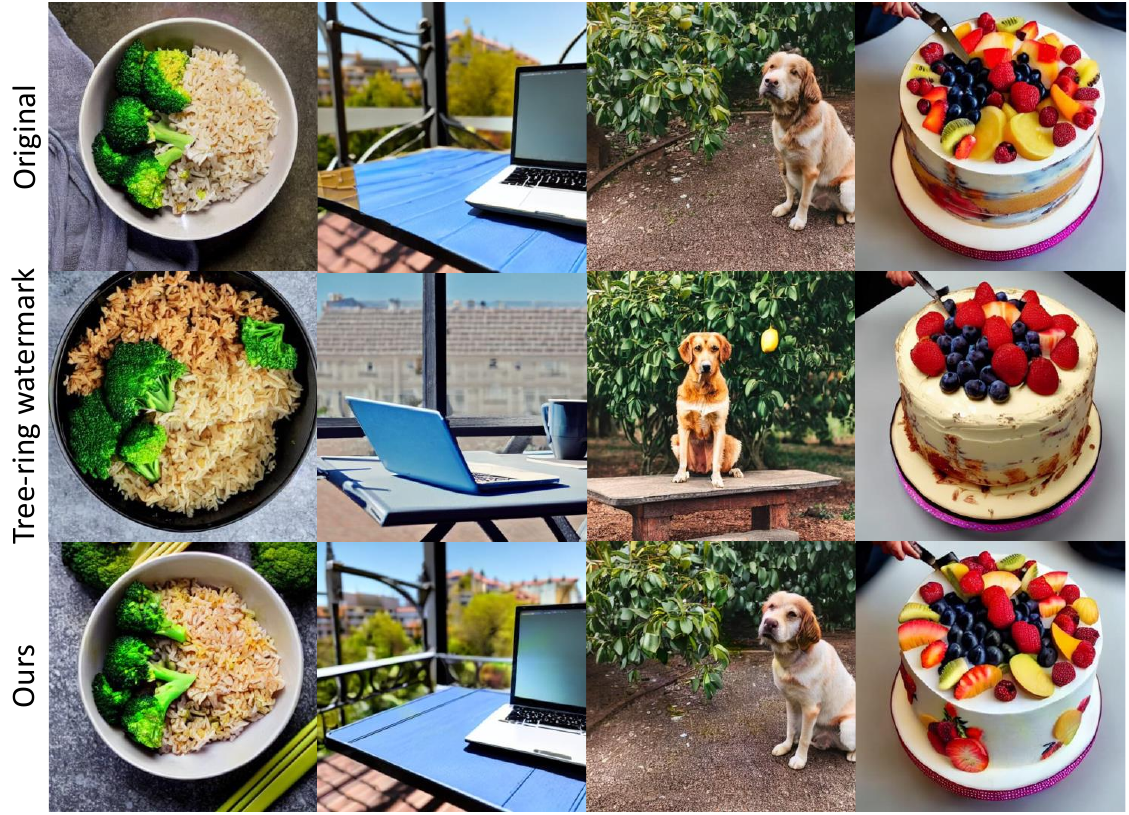}}
\caption{A comparison between the Tree-ring watermark and our proposed \texttt{\method}. The image is generated under the same diffusion configurations and the same random seed.}
\label{fig:result_comparison}
\vskip -0.1in
\end{figure}

Table \ref{tab:main_res} presents the comparison results of our method with other baseline methods. For evaluation metrics for fidelity, we adopt the Fréchet Inception Distance (FID)~\cite{heusel2017gans} calculated on the COCO2017 validation set, which comprises 5,000 images, to assess image quality.
Furthermore, we also leverage DreamSim~\cite{fu2023dreamsim}, a method that gauges the similarity between images, offering results more in line with human judgment compared to CLIP~\cite{radford2021learning} and DINO~\cite{caron2021emerging}. We include this metric because it better represents the similarity in semantics and layout than FID.

We categorize methods by their timing of watermark application---post-diffusion or during diffusion. Post-diffusion methods, applying watermarks after image generation, produce minimal pixel-level differences, reflected in a low DreamSim. In contrast, methods that add watermarks during diffusion experience an amplification of watermark differences due to the iterative denoising process, leading to relatively large changes in the final generated output. Importantly, this does not imply a significant decrease in generation quality, as indicated by only a negligible rise in FID (see Table~\ref{tab:main_res}).

In Figure~\ref{fig:result_comparison}, we also provide some visual examples of watermarked images generated by the tree-ring watermark \cite{wen2023tree} and our \Name.
It can be observed that our results and the original images share very similar layouts and consistency in the main content, despite some differences in detail.
More visual results can be found in the Appendix \ref{app:visual_results}.

\subsection{Robustness}
\subsubsection{Robustness Against Distortions}

\begin{table*}[t]
\caption{The comparison of different methods under different distortion settings. Our method demonstrates the best performance on average. The best results of each metric have been emphasized.}
\label{tab:distortion tab}
\vskip 0.1in
\begin{center}
\begin{small}
\begin{sc}
\setlength\tabcolsep{3.3pt}
\begin{tabular}{@{}lcccccccc@{}}
\toprule
\multirow{2}{*}{Methods} & \multicolumn{8}{c}{Distortions}                                                                                \\ \cmidrule(l){2-9} 
                         & ColorJitter  & Crop\&resize   & Blur          & Gaussian Noise & JPEG          & Denoising     & Denoising-v2 & Average  \\ \midrule
\textit{Bit Accuracy (\%)$\uparrow$}    & & & & & & & &      \\
DwtDctSvd & 88.60 & 49.33 & \textbf{99.07} & 69.91  & 84.37 & 53.12 & 49.44 & 70.55 \\
RivaGAN & 95.84 & \textbf{98.15} & 98.56 & 91.18 & 92.25 & 58.81 & 54.56 & 84.19\\
StableSig. &\textbf{96.28} & 97.39 & 90.55 & 71.78 & 85.94 & 48.58 & 48.52 & 77.01 \\
Ours & 93.38 & 91.44 & 95.85  & \textbf{93.00} & \textbf{94.92} & \textbf{87.58} & \textbf{86.83} & \textbf{91.86}\\ \midrule
\textit{TPR (FPR=$10^{-6}$)$\uparrow$}   & & & & & & & & \\
DwtDctSvd & 0.725 & 0.003 & \textbf{1.00} & 0.021 & 0.732 & 0.009 & 0.002 & 0.356 \\
RivaGAN & 0.923 & 0.957 & 0.946 & 0.707 & 0.856 & 0.002 & 0.007 & 0.630 \\
StableSig.& 0.984 & \textbf{0.988} & 0.903 & 0.347 & 0.833 & 0.002 & 0.00 & 0.580 \\
Tree-ring & \textbf{1.00} & 0.140 & 0.968 & 0.619 & 0.946 & \textbf{1.00} & \textbf{1.00} & 0.810 \\
Ours & 0.941 & 0.919 & 0.994 & \textbf{0.958} & \textbf{0.998} & 0.780 & 0.754 & \textbf{0.906}\\ \bottomrule
\end{tabular}
\end{sc}
\end{small}
\end{center}
\vskip -0.2in
\end{table*}

We evaluated the performance of our method under default settings and various image distortion conditions. For customized SD models, we downloaded 25 checkpoints from Civitai, and generated 100 images for each checkpoint, calculating the average results for evaluation. Details of 25 checkpoints can be seen in the Appendix \ref{app:evalmodels}.
We utilized parti-prompts~\cite{yu2022scaling}, from which we removed prompts categorized as ``basic" as they are too short. Considering our intention to test 25 models, we randomly selected 100 prompts. For evaluating stable signature, we used these 100 prompts with 10 random seeds to generate 1,000 images. Similarly, for traditional image watermark methods, we sampled 1,000 images from the clean SD model, added image watermarks, and calculated accuracy. We referred to the design from \cite{fernandez2023stable} and used true positive rate (TPR), controlling false positive rate (FPR) at $10^{-6}$ as an evaluation metric. Additional explanations about these metrics can be found in Appendix \ref{app:compdetail}.

Table \ref{tab:distortion tab} compares \Name and other methods under distortions. Among these transformations, the ``Denoising'' leverages the diffusion model itself. It first adds noise and then uses a clean diffusion model for denoising, allowing for the erasing of the watermark \cite{zhao2023generative}. We categorize it as a type of distortion because it's already a basic operation for various AI art tools. Detailed distortion settings can be found in Appendix \ref{app:distortion detail}.

Our method demonstrates strong resilience to various distortions, achieving the best results against ``JPEG'', ``Noise'' and ``Denoising'' while obtaining comparable robustness in other cases. Notably, the proposed \Name is currently the only solution for the white-box protection scenario, making it more reliable in practical scenarios.

\subsubsection{Robustness for Sampling Configurations}

\begin{table}[htb]
\caption{Extraction bit accuracy under different diffusion configurations. Default test settings are colored by gray cells. ``ConsistencyDec." is an abbreviation for ``ConsistencyDeccoder". }
\label{tab:Sampling_config}
\vskip 0.1in
\begin{center}
\begin{small}
\begin{sc}
\setlength\tabcolsep{2pt}
\begin{tabular}{@{}llcc@{}}
\toprule
\multicolumn{2}{l}{Configurations} & Bit Acc.(\%)$\uparrow$ & DreamSim$\downarrow$ \\ \midrule
\multirow{6}{*}{Sampler}       & DDIM       & 95.72 & 0.201\\
                               & DPM-S      & 95.74 & 0.201\\
                               & \cellcolor{lightgray}DPM-M & 95.79 & 0.201\\
                               & Euler      & 95.75 & 0.201\\
                               & Heun       & 95.76 & 0.201\\
                               & UniPC      & 95.63 & 0.200\\ \midrule
\multirow{4}{*}{Steps}         & 15         & 95.64 & 0.207\\
                               & \cellcolor{lightgray}25 & 95.79 & 0.201\\
                               & 50         & 95.20 & 0.202\\
                               & 100        & 94.98 & 0.203\\ \midrule
\multirow{3}{*}{CFG}           & 5.0        & 96.62 & 0.195\\
                               & \cellcolor{lightgray}7.5 & 95.79 & 0.201\\
                               & 10.0       & 94.55 & 0.209\\ \midrule
\multirow{3}{*}{VAE}        & sd-vae-ft-mse & 95.85 & 0.204 \\
                            & ClearVAE      & 95.80 & 0.208 \\
                            & ConsistencyDec.  & 95.32 & 0.206 \\ \bottomrule
\end{tabular}
\end{sc}
\end{small}
\end{center}
\vskip -0.1in
\end{table}

\noindent\textbf{Regular Sampling Configurations.} We explored the impact of various samplers, sampling steps, and Classifier-Free Guidance (CFG) scales \cite{ho2022classifier}  on watermark extraction in the denoising process in Table~\ref{tab:Sampling_config}.
For samplers, we evaluate on DDIM \cite{song2020denoising}, DPM-solver singlesteps, DPM-solver multisteps \cite{lu2022dpm}, Euler and Heun Sampler \cite{karras2022elucidating}, and Uni-PC samplers \cite{zhao2023unipc}. Despite different samplers, the watermark extraction rate remained largely unaffected. 
Besides, Table~\ref{tab:Sampling_config} also shows that \Name exhibits good extraction accuracy facing different sampling step settings and CFG scales.

\begin{table}[t]
\caption{Extraction bit accuracy for different output image sizes.}
\label{tab:size}
\vskip 0.1in
\begin{center}
\begin{small}
\begin{sc}
\setlength\tabcolsep{6pt}
\begin{tabular}{@{}ll|ccccc@{}}
\toprule
\multicolumn{2}{l}{\multirow{2}{*}{Bit Acc.(\%)$\uparrow$}} & \multicolumn{5}{c}{Width}             \\ \cmidrule(l){3-7} 
\multicolumn{2}{l}{}                                        & 512   & 576   & 640   & 704   & 768   \\ \midrule
\multirow{5}{*}{Height}                 & 512                & 92.79 & 91.84 & 91.82 & 91.48 & 90.15 \\
                                        & 576                & 93.38 & 91.63 & 91.50 & 91.48 & 90.63 \\
                                        & 640                & 93.48 & 92.94 & 91.88 & 91.02 & 91.02 \\
                                        & 704                & 92.85 & 92.75 & 92.29 & 92.33 & 88.56 \\
                                        & 768                & 91.33 & 90.90 & 88.87 & 89.56 & 86.04 \\ \bottomrule
\end{tabular}
\end{sc}
\end{small}
\end{center}
\vskip -0.1in
\end{table}


Furthermore, Stable Diffusion can effectively produce images in multiple sizes. Since our watermark is trained on a dataset of $512\times512$, there is a decrease in watermark extraction accuracy when sampled at larger sizes. To address this, we designed a special augmentation during the latent watermark pre-training stage, as well as conducted decoder-only fine-tuning after PPFT. Details can be found in the Appendix~\ref{app:robenhance largesize}. Table~\ref{tab:size} demonstrates the results of our method's extraction accuracy at different sampling sizes. Despite the increase in size, extraction accuracy decreases but remains practical.

\noindent\textbf{Different VAE Decoder.} For Stable Diffusion, there are several VAE decoders in the wild to choose from. Users can select different VAE decoders to transform the latent into images. We gather the VAE decoder from 3 sources: Improved decoder \citealp{sd-vae-ft-mse} released by StabilityAI, the community fine-tuned popular VAE decoder \citealp{ClearVAE}, and the ConsistencyDecoderVAE introduced in by OpenAI \cite{BetkerImprovingIG}. Our method demonstrates robustness across these variations (see Table \ref{tab:Sampling_config}); this is because our watermark exists within the U-Net. As long as the latent space for U-Net and VAE remains consistent, our watermark will appear in the final generated images.

\noindent\textbf{With ControlNet and LoRA Add-on.} In addition, we have also tested the accuracy of watermark extraction in images generated by the watermarked model when other LoRA or ControlNet \cite{zhang2023adding} is added. Our method demonstrates good robustness. For more details, please refer to the Appendix \ref{app:more ACC results}.

\noindent\textbf{Fine-tuning Attack.} From the attacker's perspective, we considered a fine-tuning attack. The experimental setup and results are in Appendix \ref{app:ft-attack}. We demonstrated that removing our watermark requires sacrificing the preservation of the model's preference.

\subsection{Ablation Studies}\label{sec:ablation}

\begin{table}[t]
\caption{An ablation study on the efficiency of PPFT, the initialization in mapping, and the impact of LoRA's rank. Default test settings are colored by gray cells.}
\label{tab:ablation}
\vskip 0.1in
\begin{center}
\begin{small}
\begin{sc}
\setlength\tabcolsep{2pt}
\begin{tabular}{lccc}
\toprule
Method & Rank & Bit Acc.(\%)$\uparrow$ & DreamSim$\downarrow$ \\ \midrule
Naive Diffusion                                             & 320                  & 48.11                                      & 0.330                         \\ \midrule
PPFT                                                        & \multicolumn{1}{l}{} & \multicolumn{1}{l}{}                       & \multicolumn{1}{l}{}          \\
\quad+ Normal init                                               & 320                  & 95.02 \color{red}(-0.77)  & 0.205                         \\
\cellcolor{lightgray}\quad+ Orthogonal init & 320 & 95.79 (+0.00) & 0.201                         \\ \midrule
PPFT                                                        & \multicolumn{1}{l}{} & \multicolumn{1}{l}{}                       & \multicolumn{1}{l}{}          \\
\quad+ Orthogonal init                                           & 128                  & 92.23 \color{red}(-3.56)  & 0.224                         \\
\quad+ Orthogonal init & \cellcolor{lightgray}320 & 95.79 (+0.00)                              & 0.201                         \\
\quad+ Orthogonal init & 512 & 96.29 \color{teal}(+0.50) & 0.192                         \\ \bottomrule
\end{tabular}
\end{sc}
\end{small}
\end{center}
\vskip -0.2in
\end{table}

In the ablation study, we set a fixed training length of 30 epochs and compared the final results.
Table \ref{tab:ablation} shows the results of the ablation study.

\noindent \textbf{Prior Preserving Fine-tuning.} We compared the differences between our proposed prior preserving fine-tuning method and Naive diffusion training. Naive diffusion training, while slowly improving accuracy, also caused significant changes to the generated results. We observed that compared to our method, the loss in naive diffusion training is two orders of magnitude larger. It is reasonable to assume that most of the loss is used to make AquaLoRA learn the distribution of the training dataset, rather than the pattern of the watermark.

\noindent \textbf{The Initialization of Mapper.} We tested the default standard normal initialization of the mapper and proposed orthogonal vector initialization of the mapper. The experimental results (Table \ref{tab:ablation}) show that the orthogonal vector initialization obtains the best performance. 

\noindent \textbf{AquaLoRA Ranks.} Through experimentation, we found that a larger rank leads to higher final extraction accuracy, but there is a diminishing marginal benefit.

\noindent \textbf{Peak Regional Variation Loss.}
PRVL Loss plays an important role during the latent watermark pre-training phase. We tested the watermark's PSNR and SSIM under three different settings: without PRVL loss, replacing PRVL loss with a similar-sized MSE loss, and using PRVL loss normally. Table~\ref{tab:prvl ablation} shows the experimental results, demonstrating that PRVL Loss yields the best results. PSNR and SSIM are not significantly expressive for local artifacts. Hence, we provide visual results to demonstrate the effectiveness of PRVL loss in Figure~\ref{fig:prvldemo2} of the Appendix.

\begin{table}[t]
\caption{PSNR and SSIM for Latent watermark pre-training stage under three loss settings.}
\label{tab:prvl ablation}
\vskip 0.1in
\begin{center}
\begin{small}
\begin{sc}
\setlength\tabcolsep{8pt}
\begin{tabular}{@{}lccc@{}}
\toprule
\textbf{} & no PRVL loss & MSE loss & PRVL loss \\ \midrule
SSIM      & 0.91                  & 0.91              & 0.92               \\
PSNR      & 29.48                 & 29.59             & 29.85              \\ \bottomrule
\end{tabular}
\end{sc}
\end{small}
\end{center}
\vskip -0.1in
\end{table}

\begin{figure}[!t]
\vskip 0.1in
\begin{center}
\centerline{\includegraphics[width=1.\columnwidth]{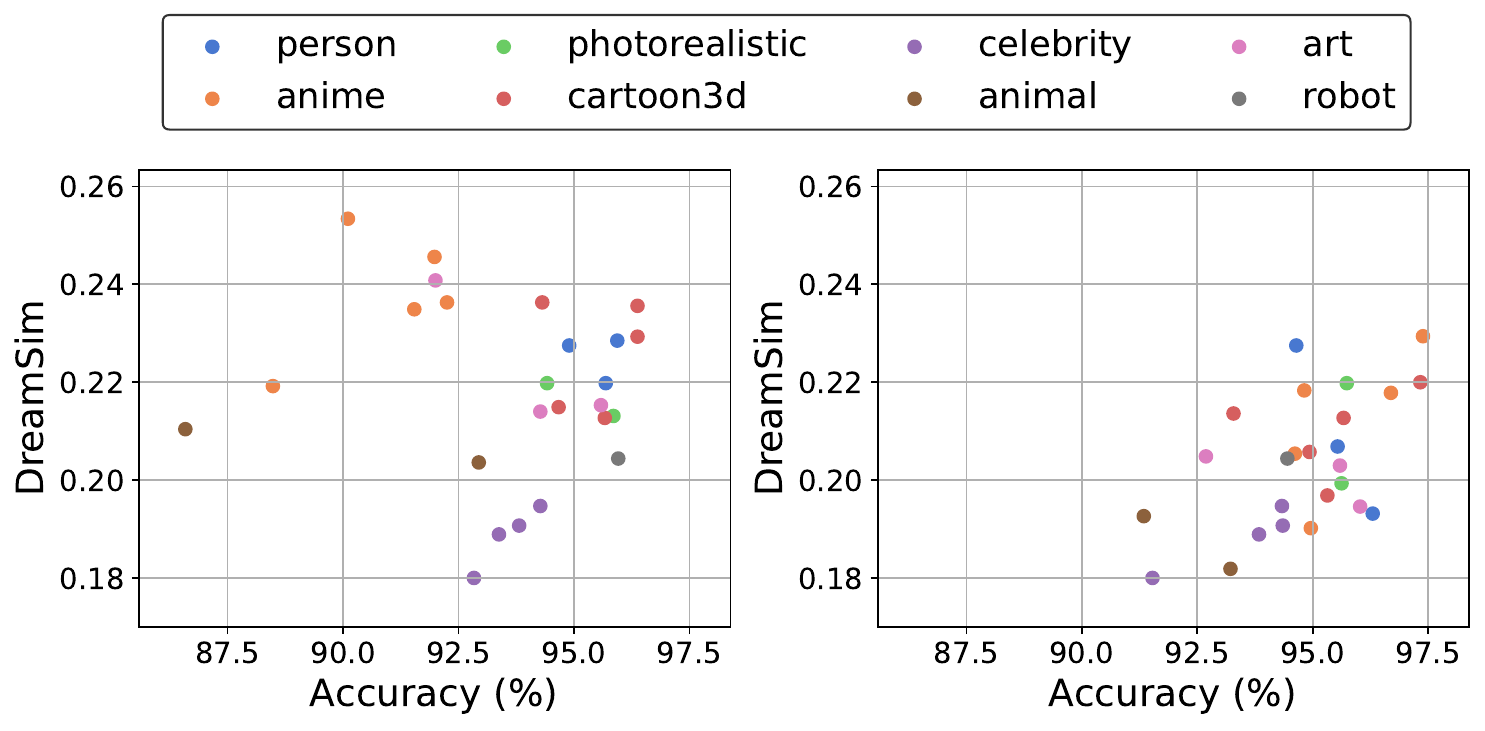}}
\caption{Ablation Study on Coarse Type Adaption. \textbf{Left}: Results for watermarked models without coarse-type adaption. \textbf{Right}: Results post fine-tuning on various coarse types.}
\label{fig:model_perform}
\end{center}
\vskip -0.4in
\end{figure}

\noindent \textbf{Coarse Type Adaption.}
As we previously mentioned, we can fine-tune our AquaLoRA on various coarse types to further enhance performance. In our experiments, we used a total of 25 downloaded checkpoints, which include two major categories: ``style'' and ``character''. Detailed information can be found in the Appendix \ref{app:evalmodels}. In Figure \ref{fig:model_perform}, the left half shows the bit accuracy and DreamSim of each model with the watermark added, without coarse-type adaption, with an average of 93.61\% and 0.219. The right half displays the results after fine-tuning, with an average of 94.81\% and 0.204.

\section{Discussion}

\textbf{Limitations.} Firstly, our method faces challenges with strong cropping and rotation due to the watermark's limited resilience in the latent watermark pre-training stage. Currently, the watermark pre-training and PPFT are decoupled. Future methods that improve watermark embedding into the latent space could replace the first stage to enhance performance.

Moreover, some advanced users might not only apply SD for text-to-image generation tasks but also engage in various editing, inpainting, and outpainting tasks. Currently, our watermark does not adequately handle these types of model usage.

Finally, when the output image size of the model increases, we have a certain degree of performance degradation. We plan to address this issue in future research.

\textbf{Conclusion.} In this work, we present \Name, an effective way for embedding watermarks into Stable Diffusion Models. Unlike previous approaches, the watermark exists within the U-Net structure, enabling protection in checkpoint-sharing scenarios (e.g., in \citealp{civitai}). By exploring the structure of LoRA, we introduce a scaling matrix that allows flexible secret modifications. Besides, we propose a prior preserving fine-tuning algorithm that embeds watermarks while ensuring minimal visual impact. Extensive evaluations across various models and experimental settings demonstrate the robustness of \Name. We analyze the limitations of our method and suggest many improvements can be made for future research. We hope our work can motivate the AI art community, moving forward into a future where creativity thrives while still being safeguarded.

\section*{Acknowledgements}

This work is supported in part by the Natural Science Foundation of China under Grants 62372423, 62121002, U20B2047, 62072421, 62206009, supported by the National Research Foundation, Singapore, and the Cyber Security Agency under its National Cybersecurity R\&D Programme (NCRP25-P04-TAICeN). It is also supported by the National Research Foundation, Singapore and Infocomm Media Development Authority under its Trust Tech Funding Initiative (No. DTC-RGC-04).

\section*{Impact Statement}

This paper underscores the necessity of implementing white-box protection for Stable Diffusion models and presents a practical solution. Our approach seeks to enhance the safeguarding of creators' interests and promote a more structured and constructive AI art community. For example, the proposed \Name could be applied by large-scale platforms like \citealp{civitai}, to protect the copyright of model sharers, which is beneficial for fostering a sharing-friendly atmosphere within the community. In addition to copyright protection, this method can be conveniently extended to track misuse and authenticate generated images.


\bibliography{example_paper}
\bibliographystyle{icml2024}

\newpage
\appendix
\onecolumn
\section{More Discussions}

\subsection{Visual Examples of Various VAE Decoders}

\begin{wraptable}[10]{r}{0.45\textwidth}
\vskip -0.15in
\caption{Quantitative comparison of visual similarity for different VAE decoders.}
\label{tab:vae_cmp}
\begin{center}
\begin{small}
\begin{sc}
\setlength\tabcolsep{4pt}
\begin{tabular}{@{}lccc@{}}
\toprule
                      & Dreamsim $\downarrow$ & SSIM $\uparrow$ & PSNR $\uparrow$  \\ \midrule
sd-vae-ft-mse         & 0.013    & 0.82 & 27.54 \\
ClearVAE              & 0.033    & 0.80 & 25.99 \\
ConsistencyDec.       & 0.030    & 0.70 & 24.68 \\
StableSig.            & 0.022    & 0.79 & 25.92 \\ \bottomrule
\end{tabular}
\end{sc}
\end{small}
\end{center}
\end{wraptable}

The customization of the SD primarily resides within the U-Net structure. Thus, altering the VAE decoder retains the model's custom content intact. We tried 4 different clean VAE decoders, namely sd-vae-ft-mse~\cite{sd-vae-ft-mse}, ClearVAE~\cite{ClearVAE}, ConsistencyDecoderVAE~\cite{BetkerImprovingIG}, and Stable Signature~\cite{fernandez2023stable} to replace the original decoder. We evaluated the sampling results by three metrics, \ie, DreamSim, PSNR, and SSIM. As shown in Table~\ref{tab:vae_cmp}, replacing the VAE decoder will not degrade the functionality of SDs. Our method selects more primary components (\ie, U-Net structure) of the SD model to embed the watermark, replacing UNet will destroy the functionality or customization of the SD.

\begin{figure}[htb]
\centerline{\includegraphics[width=0.8\textwidth]{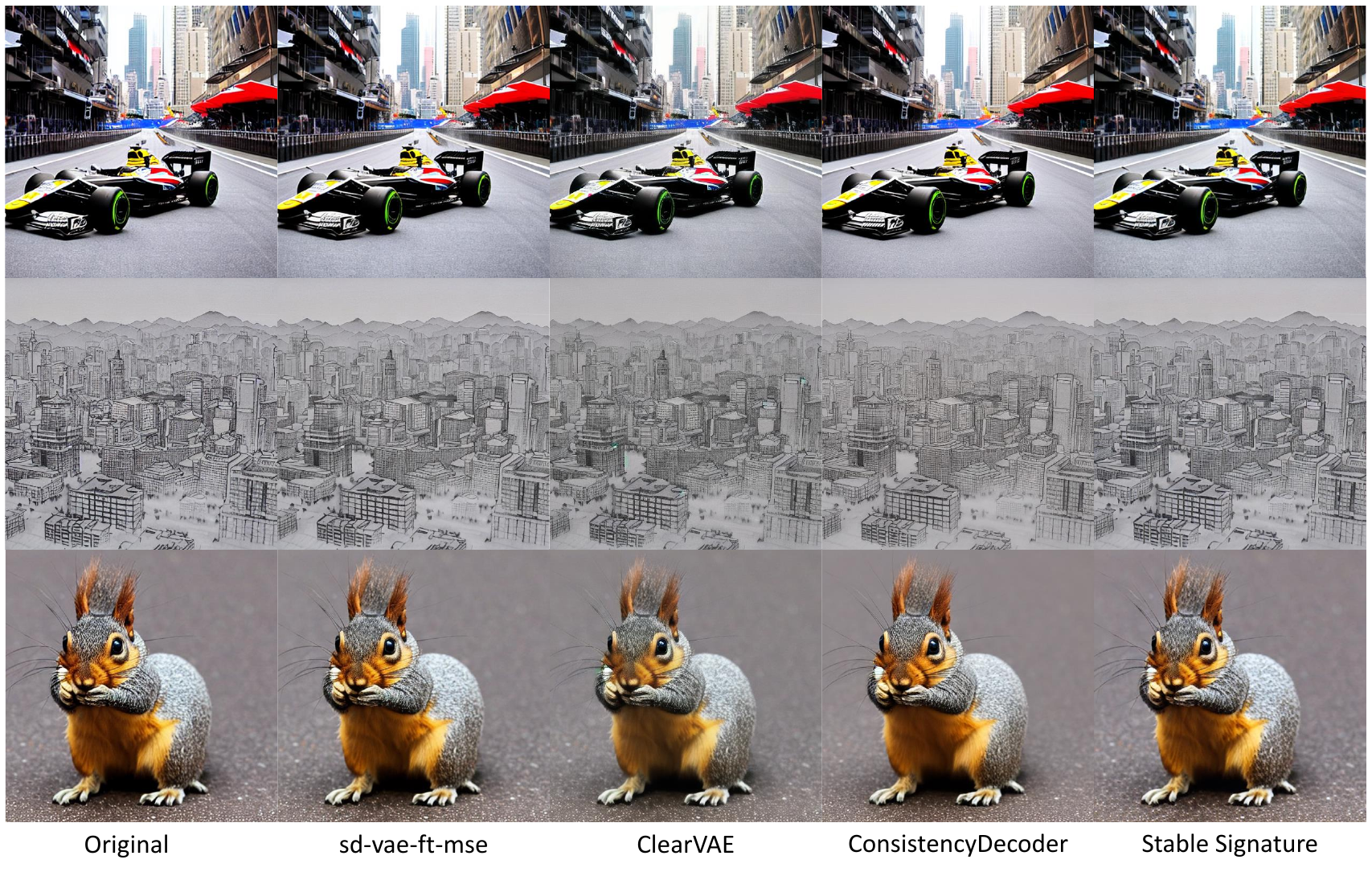}}
\caption{Representative visual examples of Stable Diffusion generated results decoded by different VAE decoders.}
\label{fig:cmpvae}
\end{figure}

Moreover, we provide a visual example (see Figure~\ref{fig:cmpvae}). It can be seen that the results generated by various VAE decoders only have very slight differences.

\subsection{Model collusion}

\begin{wraptable}[10]{r}{0.4\textwidth}
\vskip -0.15in
\caption{Expectation of extracted bits. The term ``Model1-0" refers to the positions where the watermark bits of the first model are 0. Similarly for other terms.}
\label{tab:model collusion}
\vskip 0.15in
\begin{center}
\begin{small}
\begin{sc}
\setlength\tabcolsep{8pt}
\begin{tabular}{@{}lcc@{}}
\toprule
Expectation & Model2-0 & Model2-1 \\ \midrule
Model1-0    & 0.04     & 0.53     \\
Model1-1    & 0.47     & 0.97     \\ \bottomrule
\end{tabular}
\end{sc}
\end{small}
\end{center}
\vskip -0.1in
\end{wraptable}

Users might deceive detection by aggregating their models to average their model weights, as in Model soups \cite{wortsman2022model}, creating a new model. Here, we merge two watermark models with different watermark bit strings at a ratio of 0.5. We discover that the bit at position $l$ output by the extractor will be 0 (respectively, 1) when the $l$-th bits of both models are 0 (respectively, 1), and the extracted bit is random when their bits disagree. Table~\ref{tab:model collusion} displays the average values (Expectation) of the extracted results from the merged model under different bit settings for Models 1 and 2, proving the aforementioned findings.

This aligns with the findings reported by \citeauthor{fernandez2023stable} which conforms to the \textit{marking assumption}. This so-called \textit{marking assumption} plays a crucial role in the literature on traitor tracing \cite{furon2012decoding,meerwald2012toward}. It's interesting even though our watermarking process was not explicitly designed for this, it still holds.

\subsection{Scaling \texttt{AquaLoRA} to More Bits}
Although we used a 48-bit watermark in our main experiment, we also tried extending our watermark to more bits. We increased the rank size to 512 and then tested watermarks of 64 bits and 100 bits. Table~\ref{tab:more bits} shows the experimental results, demonstrating that our method can successfully scale up to 64 bits with little performance loss. For 100 bits, there is a moderate decrease in watermark performance, which we leave as a topic for future study.

\begin{table}[ht]
\caption{Performance of our method on 64-bit and 100-bit settings. Adversarial here refers to the average performance of many different distortions.}
\label{tab:more bits}
\vskip 0.1in
\begin{center}
\begin{small}
\begin{sc}
\setlength\tabcolsep{8pt}
\begin{tabular}{@{}lcccc@{}}
\toprule
\multirow{2}{*}{\begin{tabular}[c]{@{}l@{}}Number\\ of Bits\end{tabular}} & \multicolumn{2}{c}{Fidelity} & \multicolumn{2}{c}{Robustness} \\ \cmidrule(l){2-3} \cmidrule(l){4-5}
 & FID $\downarrow$    & DreamSim $\downarrow$   & Bit Acc. $\uparrow$ & Bit Acc.(Adversarial) $\uparrow$ \\ \midrule
64-bit & 24.53 & 0.229 & 94.47                          & 88.36 \\
100-bit & 24.72 & 0.238 & 90.11                          & 83.45 \\ \bottomrule
\end{tabular}
\end{sc}
\end{small}
\end{center}
\vskip -0.1in
\end{table}

\subsection{Computational and Time Complexity for Training and Inference}

The training phase is divided into latent watermark pre-training and prior-preserving fine-tuning. During the latent watermark pre-training phase, we train for 40 epochs, approximately 80k steps, costing 40 GPU hours on a single A6000 40G. In PPFT, we train for 30 epochs, about 30k steps, costing 15 GPU hours on an A6000 40G. This is acceptable for any normal-sized academic laboratory. It's important to note that for all customized models, we only need to pre-train a coarse-type quantity of AquaLoRA.

During the inference phase, since LoRA has been integrated into the model weights, \textbf{there is essentially 0 overhead}. At this stage, our overhead is lower than that of post-diffusion or image watermarking methods.

\subsection{Inherent Shortcomings for Cover-agnostic Watermarks}
Cover-agnostic watermark has inherent weaknesses. Consider if the attacker averages many latent vectors, he will estimate the watermark signal $\Delta z_w$. However, in practice, it requires collecting a large number of in-distribution unwatermarked samples, which remains challenging. Furthermore, the above attack can be mitigated by the following measures:
\begin{enumerate}
    \item Dynamic watermarks: Regularly update the watermark pattern or parameters, making it difficult for attackers to track and analyze the watermark over time.
    \item Apply watermarks only to significant content, reducing the number of samples available for attackers to analyze.
\end{enumerate}

\section{Details in Latent Watermark Pre-training}

\subsection{Network Architecture of Secret Encoder and Training Strategy} \label{app:secret encoder_train strategy}

\begin{wrapfigure}[10]{r}{0.4\textwidth}
\begin{center}
\vskip -0.3in
\centerline{\includegraphics[width=0.4\columnwidth]{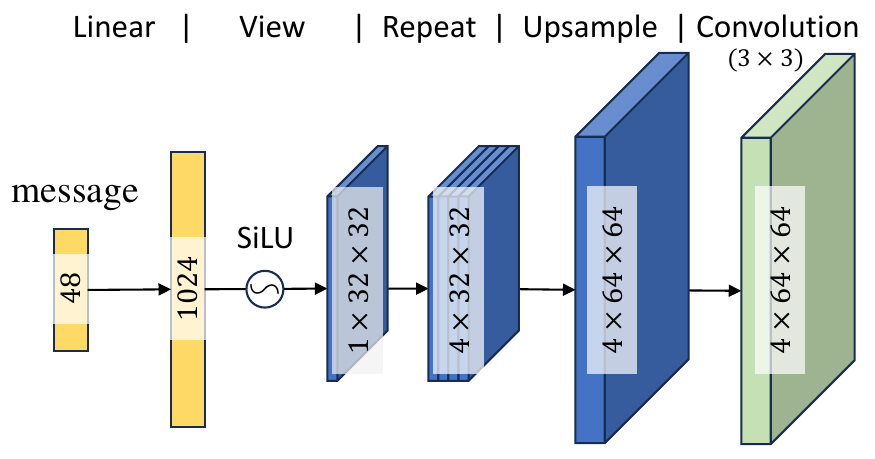}}
\vskip -0.1in
\caption{Network architecture for latent watermark encoder.}
\label{fig:wmencoderarch}
\end{center}
\end{wrapfigure} 

In the design of the secret encoder, we have drawn inspiration from the model structure of RoSteALS \cite{bui2023rosteals}. We changed the resolution from 16 to 32 to enhance the watermark's robustness against cropping operations. We removed the final zero convolution because it had a minor impact on the training process. Instead, it tended to slow down the training speed.

Regarding our training strategy, in the initial phase of training, we retained only the $\mathcal{L}_{\mathrm{BCE}}$ and did not use natural image datasets. Instead, we directly used the output of the secret encoder as the input for the VAE decoder for training, demanding accurate watermark extraction. Only when the average loss over 10 iterations fell below 0.1, did we introduce the natural image dataset, MSCOCO2017 trainset, into the training process. From that point on, we trained using the complete loss as mentioned in the main body. Without this training strategy, we find that the loss can hardly decrease.

\subsection{Details of the Peak Regional Variation Loss} \label{app:PRVL detail}

\begin{figure}[htbp]
\centering
\begin{minipage}[t]{0.43\textwidth}
\centering
\includegraphics[width=\textwidth]{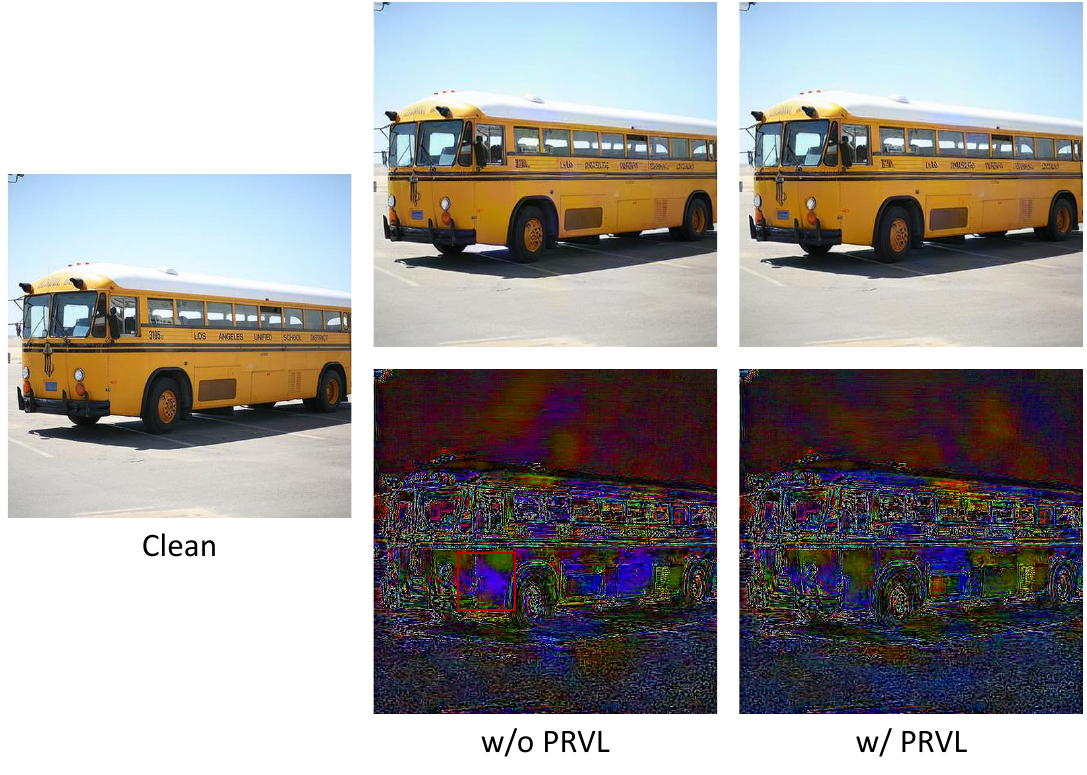}
\caption{Representative examples of our latent watermark with and without PRVL loss. The residual is amplified 10× for visualization.}
\label{fig:prvldemo}
\end{minipage}
\hfill
\begin{minipage}[t]{0.55\textwidth}
\centering
\includegraphics[width=\textwidth]{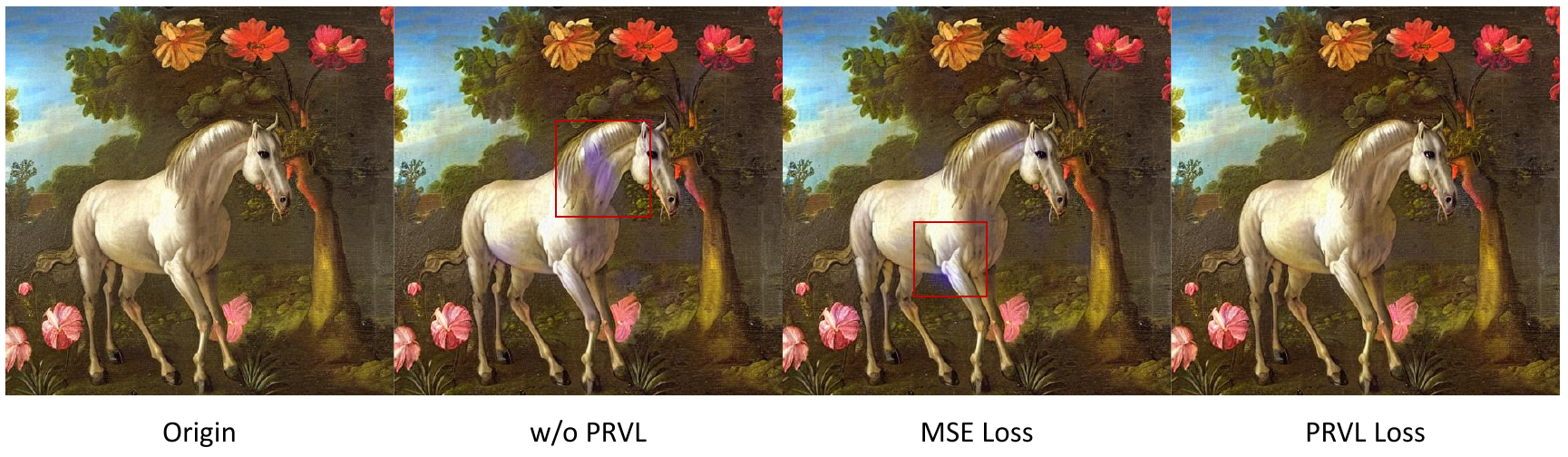}
\caption{More ablation on PRVL Loss. We tested three different settings: simply removing PRVL Loss from our training scheme, replacing PRVL Loss with a similarly sized MSE Loss, and the original training scheme. It can be observed that the watermark results from the first two settings exhibit artifacts.}
\label{fig:prvldemo2}
\end{minipage}
\end{figure}

During the training process, we found that areas with particularly strong signal features severely affect the visual quality (see Figure \ref{fig:prvldemo} w/o PRVL). Although increasing the scale of LPIPS loss can suppress this effect, it also leads to overall suppression of the watermark in invisible areas, making it difficult to improve extraction accuracy. Therefore, we specifically designed the Peak Regional Variation Loss (PRVL) which is engineered to focus on the maximal discrepancy within a predefined window or region. This is achieved by computing the absolute difference between the corresponding pixels of the two images and aggregating these differences across all color channels to form a combined variation map. The loss then centers on the region exhibiting the peak variation, identified via a convolution operation with a uniform kernel over the combined map. This approach ensures that PRVL is not unduly influenced by widespread, low-level variations but rather emphasizes areas of maximal discrepancy. Specifically, this loss can be formalized as follows:
\begin{align}
V(x, y) &= \frac{1}{3} \sum_{c=1}^{3} \left| I^{o}_{c}(x, y) - I^{w}_{c}(x, y) \right| \\
\mathcal{L}_{\mathrm{PRVL}} &= \max_{x, y} (V * K)(x, y).
\end{align}

$V(x,y)$ is a 2D tensor representing the average variation at each pixel. $K$ represents a uniform convolution kernel used to aggregate localized variations over a defined window size. $I_{c}(x, y)$ represent an image, $c$ is the corresponding channel and $x,y$ stand for position of a pixel.
We show the results with and without PRVL loss in Figure \ref{fig:prvldemo}.

\section{Details of the Distortion Settings} 

\subsection{Distortion Simulation Layer in Training Stage}\label{app:distortion layer}
In our training, we employed JPEG, crop and resize, Gaussian blur, Gaussian noise, and color jitter as our distortion simulation layers. For JPEG, we utilized the simulation layer from HiDDeN\cite{zhu2018hidden} for JPEG distortion. For the other distortions, we used \texttt{RandomCrop} and \texttt{Resize} from \texttt{torchvision}, initially randomly altering the width and height within the range of $[256, 512]$, and then resizing back to $512 \times 512$. \texttt{RandomGaussianBlur}, \texttt{RandomGaussianNoise}, and \texttt{ColorJiggle} are from the \texttt{Kornia} library. For \texttt{RandomGaussianBlur}, we randomly chose a kernel size from $[3,9]$ with an intensity selection of $(0,2]$. For \texttt{RandomGaussianNoise}, we set the mean to $0$ and the variance to $10$. For \texttt{ColorJiggle}, we adjusted the brightness in $(0.8,1.25)$, contrast in $(0.8,1.25)$, saturation in $(0.8,1.25)$, and hue in $(-0.2,0.2)$.

Additionally, to enhance the watermark robustness on images sampled at larger sizes, we propose a new augmentation, which we discuss in detail in Appendix~\ref{app:robenhance largesize}.

\subsection{Distortion in Evaluation Stage}\label{app:distortion detail}
During the testing phase, we applied lossy compression to the images using JPEG, employing the \texttt{PIL} library with a quality setting of 50. For cropping, we used a random crop of 80\%. Gaussian blur, Gaussian noise, and color jitter were applied using functions from the \texttt{Kornia} library. In Gaussian blur, we used a kernel size of $3 \times 3$ with an intensity of $4$. For Gaussian noise, we set the mean to $0$ and the variance to $0.1$ (image is normalized into $[0,1]$). In color jitter, we sampled brightness from $(0.9,1.1)$, contrast from $(0.9,1.1)$, saturation from $(0.9,1.1)$, and hue from $(-0.1,0.1)$. 

For denoising, we used Stable Diffusion v1.5, with a noise strength of 0.1, meaning the added noise was equivalent to the noise intensity of 100 steps in the DDPM forward process (Stable Diffusion has a total of 1000 timesteps). For denoising-v2, we employed Stable Diffusion v2.1, with a noise strength of 0.2. Existing watermarking methods often add watermarks at the pixel level, whereas current generative models compress or regenerate images at the semantic level, significantly leading to the loss of watermark information. This was experimentally demonstrated in \cite{zhao2023invisible}, showing that compression can significantly eliminate image watermarks. Differently, our watermarks are added to the latent space, where they are more prominent.

\begin{figure}[ht]
\centerline{\includegraphics[width=0.85\textwidth]{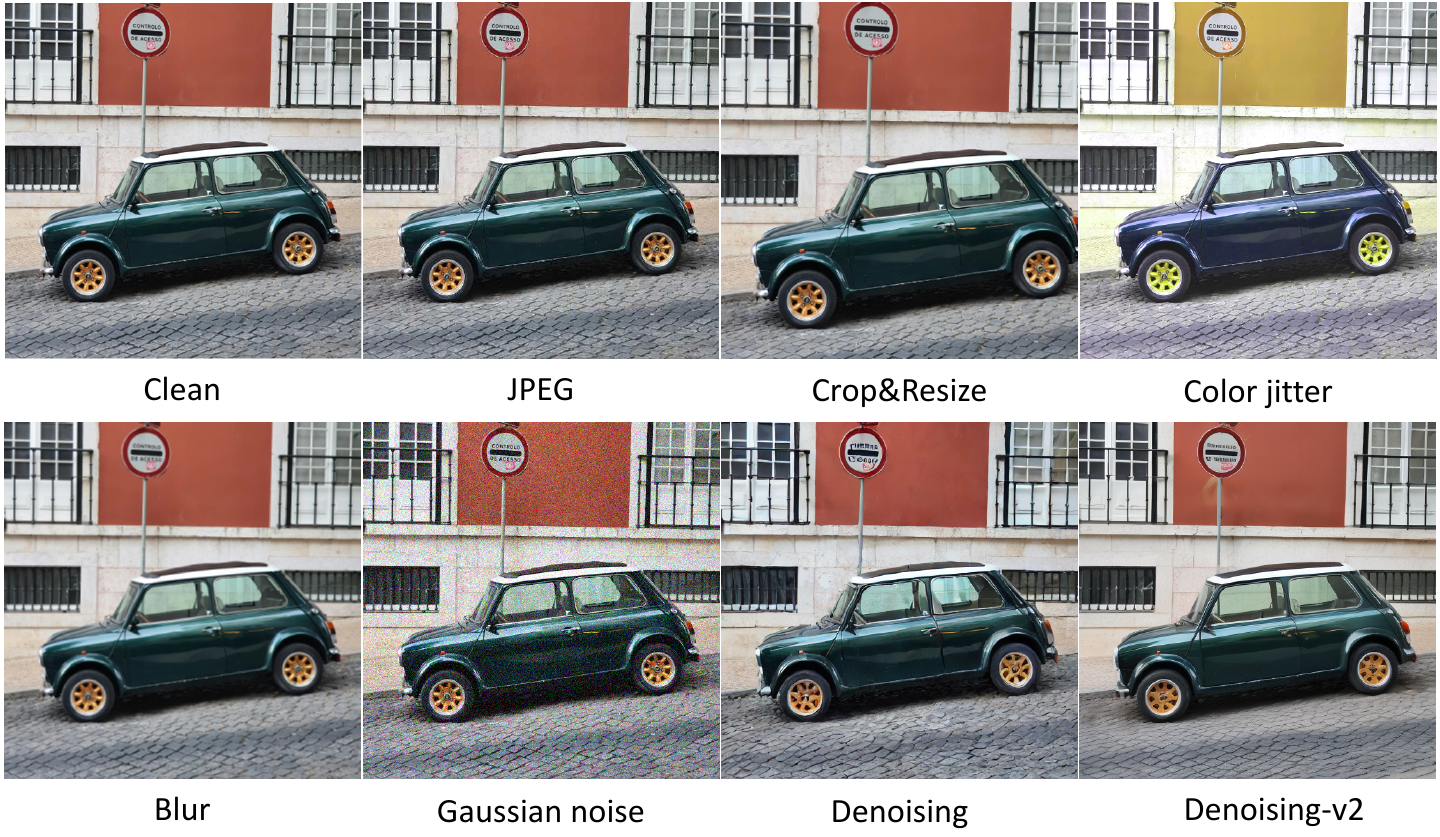}}
\caption{Demonstration of the various visual results of distortions that were used during the evaluation}
\label{fig:ds_demo}
\vskip -0.2in
\end{figure}

Figure \ref{fig:ds_demo} shows the various visual results of distortions that we used during the evaluation.

\section{Details of Robustness Enhancement on Larger Sampling Sizes}\label{app:robenhance largesize}
We designed an augmentation to enhance the watermark extraction capability of our method at larger sampling sizes, and after the PPFT stage, freeze all remaining weights to conduct additional fine-tuning on the decoder.

\noindent \textbf{Augmentation.} We considered the following question: How does a watermark pattern behave under larger sampling sizes after being trained on $512\times512$ watermarks? We observed that the patterns stayed almost the same as the original size watermark pattern near the four corners. Based on this, we designed the following process during the latent watermark pre-training phase, allowing the secret encoder and decoder to optimize this observation. Specifically, we divided the $4\times64\times64$ watermark pattern \(\Delta z_w\) into four $4\times32\times32$ patches, then resized the $4\times64\times64$ image latent \(z_o\) to 1 to 1.5 times its original size to simulate a larger image. Afterward, we overlaid the four patches onto the corners of the image latent to produce the watermarked image latent \(z_w\) and then resized \(z_w\) back to the original size. The VAE decoder decodes this to yield the watermarked image \(I_w\), completing the watermarking process.

\noindent \textbf{Fine-tuning.} 
In addition to adding augmentation, we incorporated a decoder-only fine-tuning step after PPFT training to enhance the extraction accuracy. Specifically, we used the Stable-Diffusion-Prompts~\cite{Gustavosta} as the prompt dataset and sampled images of varying sizes, ranging from $512\times512$ to $768\times768$, using different secret messages. Subsequently, we utilized the decoder to extract the secret message from the images and calculated the BCE Loss. During the training process, we froze all weights except for those of the decoder, focusing on enhancing the decoder's capabilities.

\section{Mathematical Proof of PPFT} \label{app:PPFT-math}

In this proof, we omit the text condition $c$ of the model. Assuming that the original distribution of the model is \(q\) with parameters \(\vartheta\), our objective is to learn the target distribution \(q'\) with model parameters \(\theta\). Formally, we define \(q'\) as a distribution that satisfies:
\begin{equation}
q(z_0) = q'(z_0 + \Delta z_w) = q'(z_0').
\label{eq:0}
\end{equation}

Assume \(z_0 \sim q\), \(z'_0 = z_0 + \Delta z_w\), \(z'_0 \sim q'\). We use \(p_{\theta}\) to represent the distribution of the target model. Starting from ELBO, it's evident that we wish for the distribution \(p_{\theta}\) of the target model to minimize $L$:
\begin{equation}
\mathbb{E}\left[-\log p_{\theta}\left(z'_{0}\right)\right] \leq \mathbb{E}_{q'}\left[-\log \frac{p_{\theta}\left(z'_{0: T}\right)}{q'\left(z'_{1: T} \mid z'_{0}\right)}\right]=\mathbb{E}_{q'}\left[-\log p\left(z'_{T}\right)-\sum_{t \geq 1} \log \frac{p_{\theta}\left(z'_{t-1} \mid z'_{t}\right)}{q'\left(z'_{t} \mid z'_{t-1}\right)}\right]=: L.
\end{equation}

Further, we can deduce:
\begin{equation}
L = \mathbb{E}_{q'}[\underbrace{D_{\mathrm{KL}}\left(q'\left(z'_{T} \mid z'_{0}\right) \| p\left(z'_{T}\right)\right)}_{L_{T}}+\sum_{t>1} \underbrace{D_{\mathrm{KL}}\left(q'\left(z'_{t-1} \mid z'_{t}, z'_{0}\right) \| p_{\theta}\left(z'_{t-1} \mid z'_{t}\right)\right)}_{L_{t-1}} \underbrace{-\log p_{\theta}\left(z'_{0} \mid z'_{1}\right)}_{L_{0}}].
\end{equation}
$L_T$ can be treated as a constant, \(L_0\) can be seen as a type of distortion and can be ignored. Only consider $L_{1:T-1}$.

Considering \(q(z_{t-1}|z_t, z_0)\), since $q$ is the distribution of model $\vartheta$, we can directly derive the mean:
\begin{equation}
\tilde\mu_t(z_t, z_0) = \mu_\vartheta(z_t)= \frac{1}{\sqrt{\alpha_t}} \left( z_t - \frac{\beta_t}{\sqrt{1 - \alpha_t}} \epsilon_\vartheta(z_t, t) \right),
\label{eq:1}
\end{equation}
and variance \(\tilde{\beta}_t := \frac{1 - \alpha_{t-1}}{1 - \alpha_t} \beta_t\).

Importantly, due to definition Equation \ref{eq:0}, we can obtain
\begin{equation}
q'\left(z'_{t-1} \mid z'_{t}, z'_{0}\right)=q\left(z_{t-1} \mid z_{t}, z_{0}\right).
\end{equation}
From this, it is known that the mean of \(q'\left(z'_{t-1} \mid z'_{t}, z'_{0}\right)\) is \(\mu_\vartheta(z_t)\).

Based on the KL divergence formula:
\[
KL(p, q) = \log\frac{\sigma_2}{\sigma_1} + \frac{\sigma_1^2 + (\mu_1 - \mu_2)^2}{2\sigma_2^2} - \frac{1}{2}
\]
The variance of \(q'\left(z'_{t-1} \mid z'_{t}, z'_{0}\right)\) is a fixed value, and the variance of \(p_{\theta}\left(z'_{t-1} \mid z'_{t}\right)\) is set to be a constant related to \(\beta\). Therefore, only the mean needs to be calculated.

We can obtain the effective computational part:
\begin{equation}
L_{t-1} = \mathbb{E}_{z'_0 \sim q'} \left[ \frac{1}{2\sigma_t^2} ||\tilde\mu_t(z'_t, z'_0) - \mu_\theta(z'_t, t)||^2 \right] + C = \mathbb{E}_{z_0 \sim q} \left[ \frac{1}{2\sigma_t^2} ||\mu_\vartheta(z_t) - \mu_\theta(z'_t, t)||^2 \right] + C.
\label{eq:2}
\end{equation}

Following DDPM, we perform parameterization:
\begin{equation}
\boldsymbol{\mu}_{\theta}\left(z'_{t}, t\right)=\tilde{\mu}_{t}\left(z'_{t}, \frac{1}{\sqrt{\bar{\alpha}_{t}}}\left(z'_{t}-\sqrt{1-\bar{\alpha}_{t}} \epsilon_{\theta}\left(z'_{t}\right)\right)\right)=\frac{1}{\sqrt{\alpha_{t}}}\left(z'_{t}-\frac{\beta_{t}}{\sqrt{1-\bar{\alpha}_{t}}} \epsilon_{\theta}\left(z'_{t}, t\right)\right).
\label{eq:3}
\end{equation}

By substituting Equation \ref{eq:1} and \ref{eq:3} into Equation \ref{eq:2}, we can derive the final loss function (omit text condition $c$):
\begin{align}
\begin{split}
\mathcal{L}_{\mathrm{PPFT}}(\theta) := & \mathbb{E}_{t, z_{0}, \boldsymbol{\epsilon}}\left[ \left\| \boldsymbol{\epsilon}_{\vartheta}\left(z_{t}, t\right) - \boldsymbol{\epsilon}_{\theta}\left( z'_{t}, t\right) \right\|^2 \right] \\
= & \mathbb{E}_{t, z_{0}, \boldsymbol{\epsilon}}\left[ \left\| \boldsymbol{\epsilon}_{\theta}\left(\sqrt{\bar{\alpha}_{t}} (z_{0} + \Delta z_w) + \sqrt{1-\bar{\alpha}_{t}} \boldsymbol{\epsilon}, t\right) - \boldsymbol{\epsilon}_{\vartheta}\left(\sqrt{\bar{\alpha}_{t}} z_{0} + \sqrt{1-\bar{\alpha}_{t}} \boldsymbol{\epsilon}, t\right) \right\|^2 \right].
\end{split}
\end{align}

\section{Details of Comparison Experiments} \label{app:compdetail}

\noindent \textbf{Bit Accuracy.}
Assuming a $k$-bit binary watermark $s \in \{0,1\}^{k}$ is injected into the target model, and the bit string extracted from the sampled generated image is $s'$, bit accuracy is defined as the ratio of the number of matching bits between $s$ and $s'$ to $k$, defined as $Acc(s, s')$.

\noindent \textbf{TPR with Controlled FPR.}
We consider all the watermark approach as a single-bit watermark, with a fixed watermark $s$. A threshold value $\tau$, which ranges from 0 to $k$, is predetermined. If the accuracy score $Acc(s, s')$ meets or exceeds the threshold $\tau$, it is concluded that the image indeed contains the watermark.

Previous research~\cite{yu2021artificial} commonly assumed that watermark bits \( s_1', \ldots, s_k' \) retrieved from clean images are random and uniformly distributed, each bit \( s_i' \) being modeled by a Bernoulli process with a success probability of 0.5. Consequently, the accuracy measure \( Acc(s, s') \) adheres to a binomial distribution characterized by the parameters \( (k, 0.5) \).
Once the distribution of \( Acc(s, s') \) is determined, the false positive rate (FPR) is defined as the probability that \( Acc(s, s') \) of a vanilla image exceeds the threshold \( \tau \). This probability can be further expressed using the regularized incomplete beta function \( B_x(a; b) \),
\begin{equation}
FPR(\tau) = P(Acc(s, s') > \tau) =  \sum_{i=\tau+1}^{k} \binom{k}{i} \frac{1}{2^k} = B_{\frac{1}{2}}(\tau + 1, k - \tau).
\end{equation}

We estimate the false positive rate (FPR) to be maintained at \(10^{-6}\), determine the respective threshold \( \tau \), and evaluate the true positive rate (TPR) using 1,000 watermarked images. Refer to Table~\ref{tab:distortion tab}, where the FPR is kept at \(10^{-6}\), our method demonstrates commendable performance in terms of bit accuracy and TPR.

\section{More Robustness Results} \label{app:more ACC results}

For the downloaded models, users have the option to add additional LoRA and ControlNet during the image generation process. Consequently, we conducted corresponding tests to assess this capability.

\subsection{Apply LoRA}

\begin{table}[ht]
\caption{Bit accuracy for watermarked models with LoRA add-on.}
\label{tab:addlora}
\vskip 0.15in
\begin{center}
\begin{small}
\begin{sc}
\setlength\tabcolsep{8pt}
\begin{tabular}{@{}lcc@{}}
\toprule
\textbf{}    & \begin{tabular}[c]{@{}c@{}}character +\\ style LoRA(Ghibli style)\end{tabular} & \begin{tabular}[c]{@{}c@{}}style +\\ character LoRA(Shadowheart)\end{tabular} \\ \midrule
Bit accuracy (\%) $\uparrow$ & 92.64                                                                                   & 93.91                                                                                  \\ \bottomrule
\end{tabular}
\end{sc}
\end{small}
\end{center}
\vskip -0.1in
\end{table}

In this experiment, we tested two scenarios: In the first scenario, we used a watermarked character model and added a LoRA related to a style. In the second scenario, we employed a watermarked style model and added a LoRA concerning a character.
For both scenarios, we fixed the chosen LoRA and randomly selected 4 models for testing, sampled 100 images each then calculated the average. Table \ref{tab:addlora} shows the results of our experiment. It can be observed that, although the addition of LoRA had some impact on performance, the accuracy remained above 92\%, demonstrating good robustness.

\subsection{Apply ControlNet}

In this experiment, we tested all types of the 1.0 version of ControlNet. For canny, depth, hed, MLSD, normal, and segmentation, we randomly selected 4 models from the style group. For openpose, we chose 4 models from the celebrity type. We sampled 100 images from each model to calculate the average bit accuracy. Table \ref{tab:addcontrolnet} displays the results of our experiment, which show that ControlNet did not impact the extraction of watermarks.

\begin{table}[ht]
\caption{Bit accuracy for watermarked models with ControlNet add-on.}
\label{tab:addcontrolnet}
\vskip 0.15in
\begin{center}
\begin{small}
\begin{sc}
\setlength\tabcolsep{8pt}
\begin{tabular}{@{}lccccccc@{}}
\toprule
\textbf{}    & Canny & Depth & Hed & MLSD & Normal & Seg & Openpose \\ \midrule
Bit accuracy (\%)$\uparrow$ & 94.21          & 95.39          & 93.21        & 95.15         & 95.87           & 95.36        & 94.31             \\ \bottomrule
\end{tabular}
\end{sc}
\end{small}
\end{center}
\vskip -0.1in
\end{table}

\subsection{Apply Textual Inversion}

We tested the impact of two types of textual inversion, style and character, on the accuracy of watermark extraction. Our tests were based on the RealCartoon3D model. As can be seen from Table~\ref{tab:addti}, our method achieves good accuracy.

\begin{table}[ht]
\caption{Bit accuracy for watermarked models with LoRA add-on.}
\label{tab:addti}
\vskip 0.15in
\begin{center}
\begin{small}
\begin{sc}
\setlength\tabcolsep{8pt}
\begin{tabular}{@{}lcc@{}}
\toprule
\textbf{}    & \begin{tabular}[c]{@{}c@{}}style Textual Inversion\\ (Monet style)\end{tabular} & \begin{tabular}[c]{@{}c@{}}character Textual Inversion\\ (Natalie)\end{tabular} \\ \midrule
Bit accuracy (\%)$\uparrow$ & 94.79                                                                                    & 93.34                                                                                    \\ \bottomrule
\end{tabular}
\end{sc}
\end{small}
\end{center}
\vskip -0.1in
\end{table}

\subsection{Apply LCM-LoRA}
LCM-LoRA\cite{luo2023lcm}, a LoRA model trained with Stable Diffusion base models using the consistency method, accelerates image generation to as few as four steps with any custom checkpoint model. We tested the impact of integrating LCM-LoRA and found an increase in extraction accuracy, with bit accuracy reaching 97.04\%. We hypothesize this is due to LCM-LoRA's generated images having less content and smoother colors compared to normal sampling, making our watermark more pronounced.

\subsection{Robustness Against Fine-tuning} \label{app:ft-attack}

We also took into account that some advanced attackers, after obtaining the model weights, would fine-tune the model in an attempt to eliminate the watermark. It is obvious that the value of customized models lies in their inclusion of unique preferences, which are incorporated during the author's training process and are usually not publicly released. Therefore, attackers can usually only use some easily accessible public datasets to fine-tune.

Thus, we conducted corresponding experiments. Specifically, we conducted a fine-tuning attack on a watermarked model (realcartoon3d\_v12). We used AquaLoRA with ranks of 320 and 512 to add watermarks, respectively. Then, we used the MSCOCO dataset to fine-tune the model, and we statistically analyzed the model's performance at different training steps.

\begin{figure}[htbp]
\centering
\begin{minipage}[t]{0.48\textwidth}
\centering
\includegraphics[width=\textwidth]{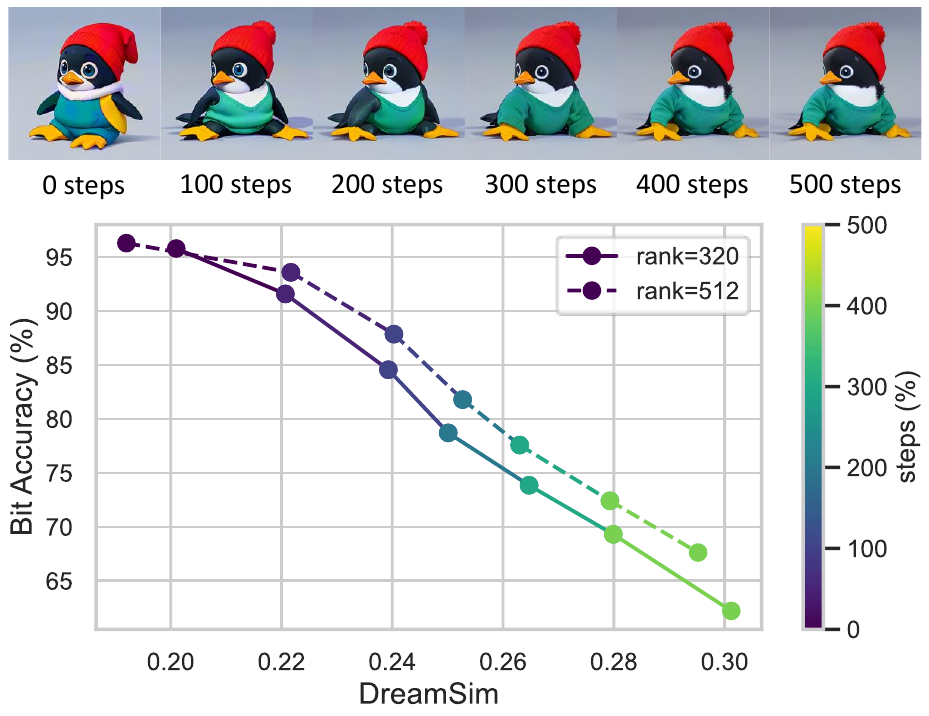}
\caption{Robustness to model fine-tuning. The image illustrates that the original style embedded in the model gradually diminishes with fine-tuning.}
\label{fig:finetune_att}
\end{minipage}
\hfill
\begin{minipage}[t]{0.48\textwidth}
\centering
\includegraphics[width=\textwidth]{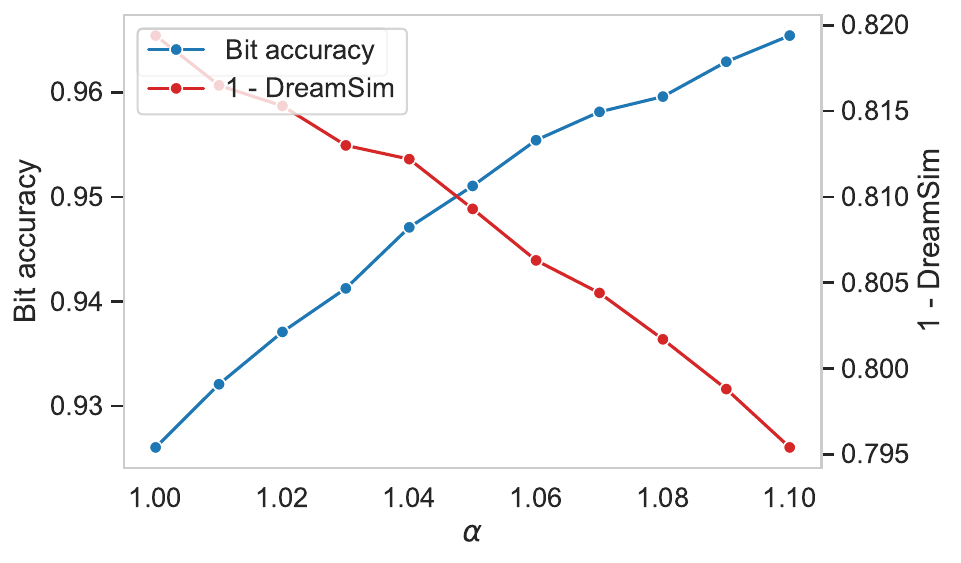}
\caption{Trade-off Between Extraction Accuracy and Image Fidelity at Different Alpha Settings.}
\label{fig:tradeoff}
\end{minipage}
\end{figure}

Our findings, as shown in Figure \ref{fig:finetune_att}, indicate that as the fine-tuning progresses, the watermark is destroyed, which also significantly compromises the integrity of the original content. Encouragingly, the figure shows that larger ranks exhibit better robustness against fine-tuning, suggesting that future research could explore using even larger ranks.

\section{More Ablation Results}

\subsection{Fidelity \& Accuracy Trade-off} \label{sec:tradeoff}

We can modify the balance between accuracy and fidelity by altering the \(\alpha\) value. This trade-off is depicted in Figure \ref{fig:tradeoff}, which shows how changes in \(\alpha\) affect both the accuracy of extraction and the fidelity of images. After considering these variations, we decided to set \(\alpha\) at 1.05.

\section{Evaluated Models and Coarse Types} \label{app:evalmodels}

See Table \ref{tab:evalmodels}.

\begin{table}[ht]
\caption{The names of the 25 models used in our experiment and their respective coarse types.}
\label{tab:evalmodels}
\vskip 0.15in
\begin{center}
\begin{small}
\begin{sc}
\setlength\tabcolsep{12pt}
\begin{tabular}{@{}lll@{}}
\toprule
Model                                  & Group     & Coarse type    \\ \midrule
counterfeitV30\_25                     & style     & anime          \\
cuteyukimixAdorable\_kemiaomiao        & style     & anime          \\
divineelegancemix\_V9                  & style     & anime          \\
meinamix\_meinaV11                     & style     & anime          \\
maturemalemix\_v14                     & style     & anime          \\
deliberate\_v11                        & style     & photorealistic \\
photon\_v1                             & style     & photorealistic \\
dreamshaper\_8                         & style     & cartoon3d      \\
juggernaut\_reborn                     & style     & cartoon3d      \\
revAnimated\_v122EOL                   & style     & cartoon3d      \\
realcartoonRealistic\_v12              & style     & cartoon3d      \\
realcartoon3d\_v12                     & style     & cartoon3d      \\
lyriel\_v16                            & style     & art            \\
neverendingDreamNED\_v122NoVaeTraining & style     & art            \\
ghostmix\_v20Novae                     & style     & art            \\
famousPeople\_caityLotz    & character & celebrity      \\
famousPeople\_evaGreen     & character & celebrity      \\
famousPeople\_sophieTurner & character & celebrity      \\
famousPeople\_aoc          & character & celebrity      \\
fenris\_v10FP16            & character & animal         \\
fluffykavkarockmergi\_v10  & character & animal         \\
chilloutmix\_NiPrunedFp32Fix           & character & person         \\
perfectdeliberate\_v5                  & character & person         \\
realisian\_v50                         & character & person         \\
robot\_V2                  & character & robot          \\ \bottomrule
\end{tabular}
\end{sc}
\end{small}
\end{center}
\vskip -0.1in
\end{table}

\section{More Visual Results} \label{app:visual_results}

\begin{figure}[ht]
\vskip 0.2in
\centerline{\includegraphics[width=0.95\textwidth]{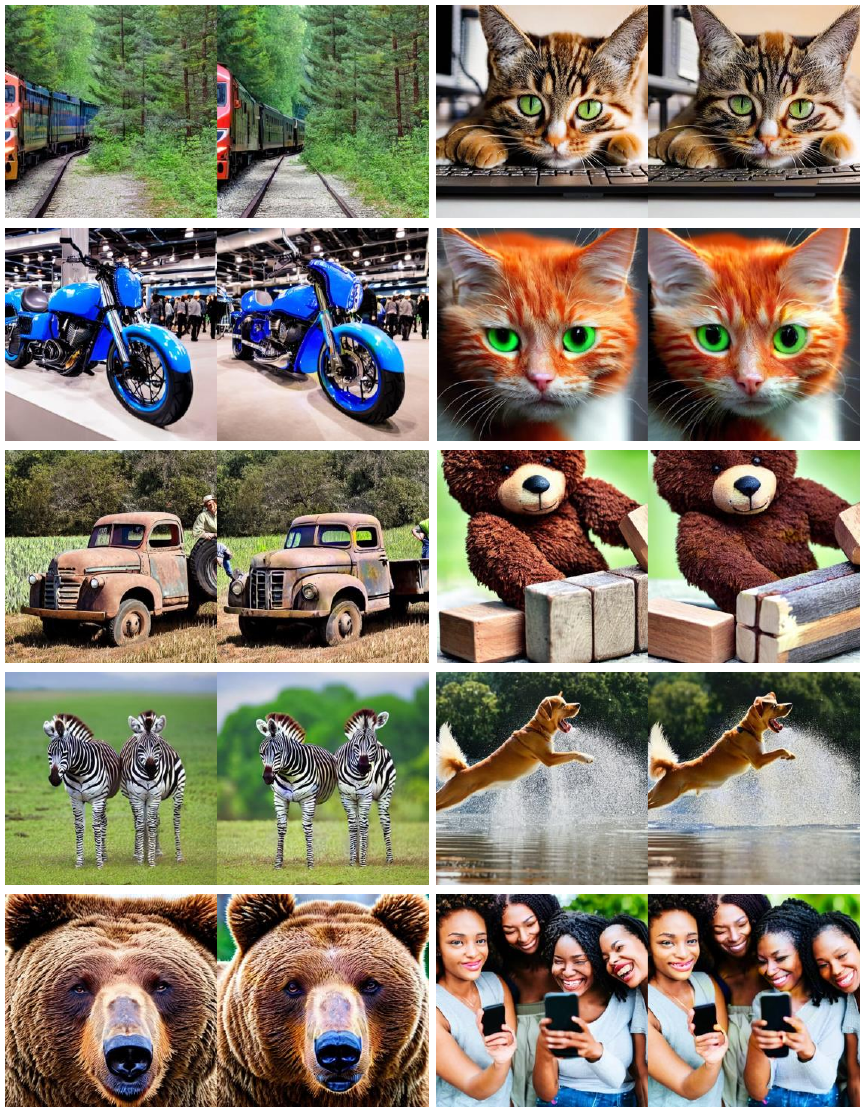}}
\caption{Comparison between images generated by the original Stable Diffusion model and the watermarked Stable Diffusion model under 
 the same diffusion configurations and random seed. \textbf{Left:} The results generated from the original model. \textbf{Right:} The results generated from the watermarked model. The results showed that the watermarked generated image is still very close to the original one.}
\label{fig:result_showcase}
\vskip -0.2in
\end{figure}

See Figure \ref{fig:result_showcase}.

\end{document}